\documentclass[fleqn,usenatbib]{mnras}

\usepackage[T1]{fontenc}

\DeclareRobustCommand{\VAN}[3]{#2}
\let\VANthebibliography\thebibliography
\def\thebibliography{\DeclareRobustCommand{\VAN}[3]{##3}\VANthebibliography}

\usepackage{graphicx}	% Including figure files
\usepackage{amsmath}	% Advanced maths commands
\usepackage{amssymb}	% Extra maths symbols
\usepackage{comment}

\usepackage{newtxtext,newtxmath}

\newcommand{\mathbi}[1]{\bmath{\it #1}}

\newcommand{\diff}{\mathrm{d}}

\newcommand{\yr}{{\rm\, yr}}

\newcommand{\Msun}{\ensuremath{\rm M_{\sun}}}
\newcommand{\Msunyr}{\ensuremath{\rm M_{\sun}\, yr^{-1}}}

\title[Accreting NS obliquity evolution]{Magnetic angle evolution in accreting neutron stars}

\author[Biryukov and Abolmasov]{
Anton Biryukov$^{1,2}$\thanks{Mailto: ant.biryukov@gmail.com} 
and Pavel Abolmasov$^{3}$
\\
$^1$Sternberg Astronomical Institute, Moscow State University, 13 Universitetsky pr., Moscow, 119234, Russia \\
$^2$Kazan Federal University, 18 Kremlyovskaya str., Kazan, 420008, Russia \\
$^3$Department of Physics and Astronomy, FI-20014 University of Turku, Finland
}

\date{Accepted XXX. Received YYY; in original form ZZZ}

\pubyear{2021}

\begin{document}
\label{firstpage}
\pagerange{\pageref{firstpage}--\pageref{lastpage}}
\maketitle

% Abstract of the paper
\begin{abstract}
%Var H: 
The rotation of a magnetised accreting neutron star (NS) in a binary system is described by its spin period and two angles: spin inclination $\alpha$ with respect to the orbital momentum and magnetic angle $\chi$ between the spin and the magnetic moment.
Magnetospheric accretion spins the NS up and adjusts its rotation axis, decreasing $\alpha$ to nearly perfect alignment. Its effect upon the magnetic angle is more subtle and relatively unstudied. In this work, we model the magnetic angle evolution of a rigid spherical accreting NS. We find that the torque spinning the NS up may affect the magnetic angle while both $\alpha$ and $\chi$ significantly deviate from zero, and the spin-up torque varies with the phase of the spin period. As the rotation axis of the NS is being aligned with the spin-up torque, the magnetic axis becomes misaligned with the rotation axis. Under favourable conditions, magnetic angle may increase by $\Delta \chi \sim 15\degr-20\degr$. This orthogonalisation may be an important factor in the evolution of millisecond pulsars, as it partially compensates the $\chi$ decrease potentially caused by pulsar torques. If the direction of the spin-up torque changes randomly with time, as in wind-fed high-mass X-ray binaries, both the rotation axis of the NS and its magnetic axis become involved in a non-linear random-walk evolution. The ultimate attractor of this process is a bimodal distribution in $\chi$ peaking at $\chi =0$ and $\chi = 90\degr$. 
\end{abstract}


%%%%%%%%%%%%%%%%%%%%%%%%%%%%%%%%%%%%%%%%%%%%%%%%%%

%%%%%%%%%%%%%%%%% BODY OF PAPER %%%%%%%%%%%%%%%%%%

\section{Introduction}

Magnetic dipole is usually a good approximation for the magnetic field of a magnetised star, at least in a certain range of radii, much larger than stellar radius and much smaller than light cylinder. 
Magnetic dipole is described by a single vector quantity (magnetic moment), that sets the direction of the magnetic axis. 
Magnetic angle $\chi$ is the angle between spin and magnetic axes of a neutron star (NS, see Figure~\ref{fig:ns_sketch}).
Though often ignored, magnetic angle is an important quantity affecting the observational properties of a NS as well as its rotational evolution. Non-zero value of magnetic angle is critical for the observations of the pulsating radiation from any type of radiating NS, accretion- or rotation-powered.

Magnetic angles, their distribution and evolution were comprehensively investigated for rotation-powered pulsars. 
Their $\chi$ were found to be nearly isotropically distributed in the broad interval from almost zero to nearly 90 degrees \citep{lyne88, rankin93b, tm98, nikitina11}. At the same time, their evolution is still a subject of discussion. 
Pulsar magnetic angle is thought to be either decreasing on the spin-down time scale \citep{phil14}, or, alternatively, increasing toward 90 degrees \citep{bes93, Novoselov2020}. In any case, for an isolated pulsar, magnetic angle as a function of time is completely determined by a single process, the pulsar losses themselves.

On the other hand, there are several populations of NSs where the rotation of the star is strongly affected by the effects of binary interaction. 
These objects include millisecond pulsars (MSP, \cite{manchester_msp_2017}) and X-ray binaries. 
The latter are normally classified according to the mass of the donor star as low- and high-mass X-ray binaries (LMXBs and HMXBs correspondingly, \citealt{bhatt91}). 
HMXB containing magnetised NSs are usually observed as X-ray pulsars \citep{CW12}.
In turn, most LXMBs have weakly magnetised accretors that does not allow to study their magnetic angles observationally (their magnetic fields are either buried by accreting matter or decay due to Joule losses in the crust, see later Section~\ref{sec:evo:phys}). 
However, accreting millisecond pulsars \citep{AMSP21}, apparently having relatively large magnetospheres, also fall into this category.
As the amount of observational data on NSs in binary systems increases, there is a growing demand for a theory describing the physics of NS rotation affected by mass accretion and additional external torques. 

Unfortunately, magnetic angle distributions in MSP and X-ray binaries are known much worse than for classical radiopulsars. Moreover, they are also affected by multiple processes, most importantly by the high-angular-momentum accretion from the binary companion.

For MSP, \citet{chen98} find an intriguing bi-modal distribution in magnetic angle. Considering a sample of 11 objects, they found that some pulsars are close to magnetic alignment, while the others are close to the orthogonal rotation. 
If the torques acting on the NS do not distinguish between the two magnetic poles and depend smoothly on magnetic angle, these are the two expected equilibrium points. 
Using a different line of reasoning, \citet{LS80} come to the same conclusion. 
However, it is hard to explain why a significant number of objects should evolve towards orthogonal rotation.  

More recently, \citet{johnson14} considered a larger sample of objects using different data (including gamma-ray data from \textit{Fermi/LAT} for 40 objects) and techniques.  
Apparently, the pulsars considered show a broad distribution in magnetic angle with no significant bi-modality, though a large number of objects still show very large (above 70\degr) magnetic angles.
This agrees with the estimates made by \citet{bogdanov08} that argue for large magnetic angles ($\chi \gtrsim 40^\circ$) in some MSP. 

MSPs are commonly understood as ``recycled'', i. e. spun-up by accretion after the end of their initial period of pulsar activity \citep{taurisheuvel}.  
This interpretation is supported by the existence of a population of accreting MSPs \citep{hessels08, AMSP21}. 
There is also at least one MSP, PSR~J1023+0038, showing pulsations both in X-ray and in radio, switching between radio-bright/X-ray-faint and radio-faint/X-ray-bright states \citep{archibald09}. 
It is tempting to interpret these two states as accretor and ejector regimes, but the recent study by \citet{veledina19} shows that most likely both states in this object are rotationally powered but require the presence of an accretion disc. 

As for X-ray pulsars, the constraints upon their magnetic field geometry are still highly uncertain and model-dependent. 
\citet{leahy90} estimated the geometry of the radiating areas of 15 X-ray pulsars using a simple non-relativistic hot spot model. 
Magnetic angles found in this study cluster around $\chi \sim 24\degr$.
Later, \citet{annala10} considering a larger sample of X-ray pulsars with relativistic models, came to a mean value of $\chi_{\rm mean} =  40\pm 4\degr$. 

In this paper, we present an attempt to trace the NS's magnetic angle evolution during
magnetospheric accretion.  In Section~\ref{sec:rot}, we derive the equations of rotational evolution including different external torques and build a phenomenological model. We show that the crucial factor for the evolution of magnetic angle is the modulation of the spin-up torque with the spin phase. 
In Section~\ref{sec:res}, we calculate the rotational evolution for four different cases aimed to represent an MSP progenitor or an X-ray pulsar. We discuss the results and consider certain extensions and implications of the model in Section~\ref{sec:disc}. Conclusions are made in Section~\ref{sec:conc}.

%__________________________________________________________________

\section{Rotational evolution of an accreting neutron star}\label{sec:rot}

\subsection{General model}
\label{sec:genmodel}

\begin{figure}
    \centering
    \includegraphics[width=1\columnwidth]{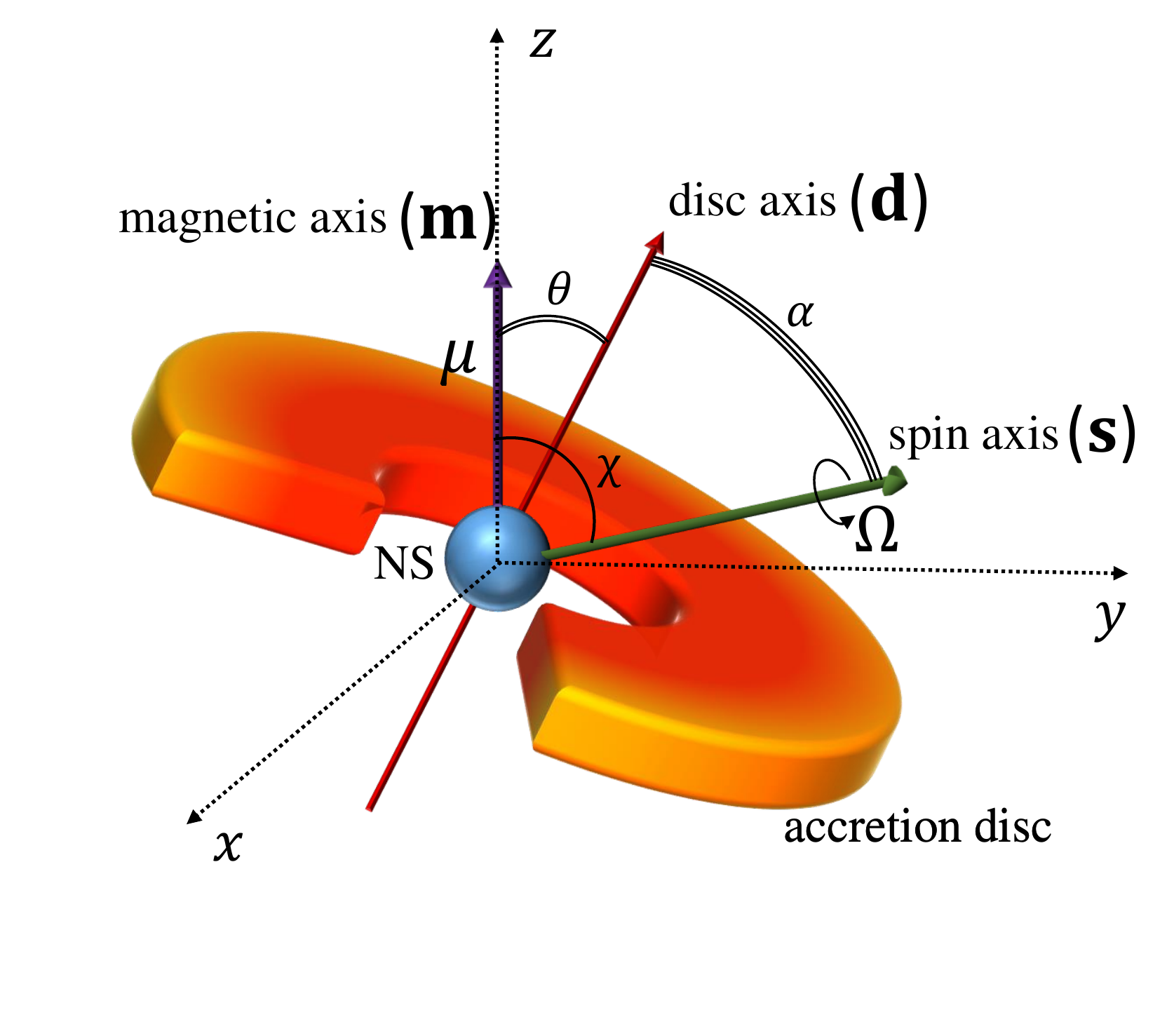}
    \caption{Main axes and angles of an accreting neutron star. % Pulsar magnetic angle is $\chi$. 
    Unit vectors along the magnetic, disc, and spin axes are $\mathbi{m}$, $\mathbi{d}$, and $\mathbi{s}$, respectively.
    The angles shown are $\chi$ (magnetic angle), $\alpha$ (rotation axis inclination with respect to the axis of the disc), and $\theta$ (instantaneous angle between $\mathbi{d}$ and $\mathbi{m}$).
    }
    \label{fig:ns_sketch}
\end{figure}

Let us consider the rotation of a solid spherical NS with variable moment of inertia $I(t)$. We will use a coordinate frame $xyz$ associated with the star as follows: axis $z$ is aligned with the magnetic moment of the star, $x$ and $y$ are perpendicular to $z$ and form a right-handed set of vectors. The orientation 
of $yz$ plane is defined by the spin $\mathbi{\Omega}$ of the NS: 
we assume that the spin axis lies  within the plane (see Figure~\ref{fig:ns_sketch}). The considered reference frame $xyz$ co-rotates with the star.  

In this frame, the rotational evolution of the NS is described in vector form by Euler's equations \citep{LL1}
\begin{equation}\label{eq:euler}
	\frac{\diff\mathbi{L}(t)}{\diff t} + \mathbi{\Omega}(t) \mathbi{\times} \mathbi{L}(t) = \mathbi{N}(t),
\end{equation}
where $\mathbi \Omega$ is the instantaneous angular velocity of the star,
\begin{equation}	\label{eq:total_momentum}
    \mathbi{L}(t) = I(t)  \mathbi{\Omega}(t)
\end{equation}
is its full angular momentum, and $\mathbi{N}$ is the external torque.

We ignore the possible oblateness of the star for the following reasons. 

Expected values of magnetic and other non-hydrostatic ellipticities of NSs are very low $\epsilon \sim 10^{-8} - 10^{-10}$. 
Their effect on the rotation of the star is restricted to very slow precession on the time scales or $\sim 10^5-10^7$ years \citep{goldreich70, melatos2000, haskell08}, that is much longer than the spin-disc alignment time scale (see Equation~\ref{eq:time_align} below). 

Hydrostatic deformation related to rotation may reach few per cent if the spin period is of the order several milliseconds, but the deformation always adds an angular momentum component aligned with $\mathbi{\Omega}$ and hence may be considered as a correction to the variable scalar moment of inertia. In Appendix~\ref{sec:app} below we consider this issue in more details.

Substituting Equation~(\ref{eq:total_momentum}) into Equation~(\ref{eq:euler}), we get an equation for the rotational evolution of a spherical NS in the most general form 
\begin{equation}
    I \dfrac{\diff\mathbi{\Omega}}{\diff \it t} = \mathbi{N} - \dfrac{\diff \it I}{\diff \it t} \mathbi{\Omega}. 
    %-\epsilon I \Omega_z \left [ \left (\dfrac{\dot I}{I} + \dfrac{\dot\Omega_z}{\Omega_z}  + \dfrac{\dot\epsilon}{\epsilon }\right ) \mathbi{ m} + \mathbi{\Omega \times m} \right],
	\label{eq:general_eqs}
\end{equation}

The  full torque $\mathbi{N}$ acting on an accreting star consists of three components
\begin{equation}
	\mathbi{N} = \mathbi{N}_{\mathrm{acc}} + \mathbi{N}_{\mathrm{mag}} + 
	\mathbi{N}_{\mathrm{psr}},
\end{equation}
where $\mathbi{N}_{\mathrm{acc}}$ describes the NS spin-up due to  accretion, 
$\mathbi{N}_{\mathrm{mag}}$ is magnetic braking caused by disc-magnetosphere interaction, and $\mathbi{N}_{\mathrm{psr}}$ corresponds to pulsar losses. 
In principle, interaction between the disc and the magnetic field also creates additional precession and warping torques (see \citealt{LS80, wang_inclined, lai11, lai2014}). 
It is easy to show, however, that if the interaction region between the magnetosphere and the disc is small, the warping torque is also small and thus we omit it in our basic simulations. Nevertheless, we will still consider the effect of the warping torque below in Section~\ref{sec:disc:warp}.

In the co-rotating $xyz$ frame we use, $\mathbi{N}_{\mathrm{acc}}$ rotates around $\mathbi{\Omega}$ with NS spin frequency $-\Omega$, while the other two components of the torque change relatively slowly retaining their direction in the frame of the star on a spin time scale.

At the same time, we expect that, if the angle $\theta$ between the
NS magnetic moment and accretion 
disc axis changes with spin phase, both the instantaneous mass accretion rate and, therefore, spin-up torque acting on the NS are modulated. Thus, we will consider the spin-up torque $\mathbi{N}_{\mathrm{acc}}$ in the form
\begin{equation}
    \mathbi{N}_{\mathrm{acc}} = N_0  f(\cos\theta)  \mathbi {d},
    \label{eq:accretion_torque}
\end{equation}
where
\begin{equation}\label{eq:accretion_torque:0}
    N_0 = \dot M \sqrt{G M_{*} r_{\mathrm{m}}},
\end{equation}
and $\mathbi d$ is the unit vector in the direction of the disc axis.
Here, $\dot{M}$ is mass accretion rate, $M_*$ is the mass of the NS, and $r_{\rm m}$ is the radius of the magnetosphere.
Vector $\mathbi{d}$ rotates around the spin axis $\mathbi{\Omega}$ in the direction opposite to the rotation of the star so that
\begin{equation}
\dot{\mathbi{d}} = -\mathbi{\Omega} \times \mathbi{d}.
\end{equation}
Geometrically
\begin{equation}
    \mathbi{d} = \left ( \begin{array}{c}
        \sin\alpha\sin\psi \\
        \sin\chi\cos\alpha - \cos\chi\sin\alpha\cos\psi \\
        \cos\chi\cos\alpha + \sin\chi\sin\alpha\cos\psi \\
    \end{array} \right ),
\end{equation}
where $\alpha$ is the angle between $\mathbi d$ and $\mathbi \Omega$, or spin inclination with respect to the disc. 
%\textbf{
Here, $\psi = -\int \Omega \diff t$ is the rotational phase equal, up to the constant of integration, to the dihedral angle between the $\mathbi{\Omega}-\mathbi{m}$ and $\mathbi{\Omega}-\mathbi{d}$ planes, and $\mathbi m = \mathbi \mu/\mu$ is the $z$-axis basis unit vector directed along the magnetic moment $\mathbi{\mu}$, so that
%}
\begin{equation}
    \mathbi{m} = \left ( \begin{array}{c}
        0 \\
        0 \\
        1 \\
    \end{array} \right ).
\end{equation}
We assume that at the time $t = 0$ all the three vectors ($\mathbi{d}$, $\mathbi{m}$, and $\mathbi{\Omega}$) lie in the same plane $yz$. 

Radius $r_{\rm m}$ in Equation~(\ref{eq:accretion_torque:0}) is the size of the magnetosphere calculated according to the classical formulae (see for instance section~6.3 of \citealt{accpower})
\begin{equation}
    r_{\mathrm{m}} = \xi\left( \dfrac{\mu^4}{ 2G M_* \dot M^2}\right)^{1/7},
\end{equation}
where $\xi \approx 0.5$ is a dimensionless constant dependent on the unknown details of the disc-magnetosphere interaction.
Dimensionless function $f(\cos\theta)$ in Equation~(\ref{eq:accretion_torque}) describes the modulation of the accretion torque as a function of the angle $\theta$ between the magnetic axis and disc axis $\mathbi{d}$. 
A phenomenological representation for this function will be introduced below in Section~\ref{sec:modulation}. 

We take the accretion term, Equation~(\ref{eq:accretion_torque}), into account only 
when the inner edge of the accretion disc is inside the co-rotation radius $r_{\mathrm{co}} = (GM_{*}/\Omega^2)^{1/3}$. 
If this condition is violated, the NS is likely to enter the so-called propeller regime when accretion is suppressed by centrifugal forces \citep{IS75}. 

As the processes responsible for the spin-down of accreting magnetised NSs are poorly constrained and probably related to the open magnetic field lines co-rotating with the star \citep{lovelace95}, we adopt the following expression for the magnetospheric spin-down torque
\begin{equation}\label{eq:nmag}
    \mathbi{N}_{\mathrm{mag}} = -\dfrac{\mu^2}{3r^3_{\mathrm{co}}}  \mathbi{s},
\end{equation}
where $\mathbi {s} = \mathbi{\Omega}/\Omega$ is a unit vector directed along the spin axis:
\begin{equation}
    \mathbi{s} = \left ( \begin{array}{c}
        0 \\
        \sin\chi \\
        \cos\chi \\
    \end{array} \right ).
\end{equation}
Though Equation~(\ref{eq:nmag}) was initially obtained in the framework of disc-magnetosphere interaction model \citep{GL77}, its general form 
also appears in the models involving magnetospheric outflows \citep{IK90,Rap2004}. 
The importance of magnetospheric outflows as a spin-down mechanism for accreting NS was confirmed by MHD simulations \citep{romanova09, ZF13, parfrey17}. 
Note the difference between the two spin-down mechanisms: pulsar torque (considered below), possibly enhanced by the increase of the number of open magnetic field lines \citep{parfrey16}, and magnetospheric ejections that we associate with $\mathbi{N}_{\rm mag}$. 

The direction of this spin-down torque has not been studied for inclined rotators, and it is difficult to predict from the basic principles. 
As the outflows are formed by centrifugal forces, we will assume that $\mathbi{N}_{\mathrm{mag}}$ is always directed along the spin axis of the neutron star. This torque depends on the size of the NS magnetopshere and can be included into the calculations only if $r_{\mathrm{m}} < r_{\mathrm{LC}} = c/\Omega$, where $r_{\mathrm{LC}}$ is the light cylinder radius. Otherwise, the magnetosphere is in the ejector regime \citep{shvartsman70, IS75} and is not affected by the possible presence of the accretion disc.

Finally, for the pulsar spin-down term, we use the results obtained by \cite{spitkovsky06} and \cite{phil14}
\begin{equation}
    \displaystyle {\mathbi {N}}_{\mathrm{psr}} = K_{\mathrm{psr}} \left \{  \left(1 + \sin^2\chi\right)\mathbi{s} + ((\mathbi{m \times s}) \mathbi{\times s}) \cos\chi\right \},
    \label{eq:pulsar_torque}
\end{equation}
where 
\begin{equation}
    K_{\mathrm{psr}} = -\dfrac{\mu^2}{r_{\mathrm{LC}}^3}
    \label{eq:pulsar_torque:basic}
\end{equation}
is the characteristic pulsar braking torque.
The first term in Equation~(\ref{eq:pulsar_torque}) describes the torque spinning the NS down while the second one is related to the evolutionary secular decrease of obliquity. Another possible component of the pulsar loss is the so-called ``anomalous torque'' \citep[e.g.][]{bz_anomal2014}. This torque
is proportional to $\mathbi \Omega \times \mathbi m$ and causes a
forced precession of the NS spin axis around its magnetic moment. 
However, as we will show in Section~\ref{sec:modulation}, the contribution of this precession term to all the evolutionary equations is exactly zero. 
Therefore, we ignore the anomalous torque in our calculations. 
%It, however, makes zero contribution to the evolution of NS's magnetic and spin-disc inclination as we will show in the next section.
% And therefore it can be omitted in our calculations. 

The more important issue, however, is the validity of
Equation~(\ref{eq:pulsar_torque}) during accretion. 
It seems clear that accreting plasma changes the structure of the magnetosphere near the polar caps and affects the distribution of magnetospheric currents and, therefore, the character of the torques acting on the star. 
However, pulsar spin-down time scale,
\begin{equation}
    \tau_{\rm psr} \sim \frac{c^3 I}{\mu^2 \Omega^2} \simeq 10^8 I_{45}   P_{\rm s}^2  \mu^{-2}_{30}\hspace{0.2cm}\yr,
    \label{eq:time_psr}
\end{equation}
is much longer than that of the spin-up from the accretion 
\begin{equation} %\label{E:res:taualign}
    \tau_\mathrm{acc} \sim I\Omega/N_0 \simeq 600 I_{45}  \dot M_1^{-6/7}  
    \left (\dfrac{M_*}{\Msun} \right)^{-9/7}  \mu^{-2/7}_{30}\yr.
    \label{eq:time_align}
\end{equation}
Here $I_{45} = I/10^{45}{\rm g\, cm^2}$, $P_{\rm s} = P / 1$s, $\mu_{30} = \mu/10^{30}{\rm\, G\, cm^3}$, 
and mass accretion rate $\dot M_1 = \dot M/\dot M_{\rm Edd,{\sun}}$ is 
normalised by the Eddington accretion rate for one Solar mass, namely 
$\dot M_{\mathrm{Edd, {\sun}}} = 4\upi G\Msun /\varkappa_{\rm T}c 
\approx 1.6\times 10^{17}{\rm \, g \,s^{-1}} 
\approx 2.5\times 10^{-9}\ \Msunyr$. 
Even for MSP $\tau_\mathrm{psr}\gtrsim 10^8\yr$ \citep{manchester_msp_2017} 
because of smaller magnetic fields $\mu_{30} \sim 10^{-4}-10^{-2}$.

Hence, pulsar torque in the form (\ref{eq:pulsar_torque}) somehow affects the NS obliquity evolution only during the ejector stage. So, we always keep the pulsar losses on during our simulation for the sake of simplicity. 

 It is also possible that the presence of an accretion disc increases the pulsar spin-down torques \citep{parfrey16}. 
We discuss this possibility in Section~\ref{sec:disc:pulsar}.

\subsection{Averaged torque}\label{sec:modulation}

In the reference frame we use, accretion torque (\ref{eq:accretion_torque}) rotates rapidly around the 
spin axis of the NS. Its projection upon the spin axis may vary with the rotational phase $\psi$, that is of primary importance for the evolution of obliquity $\chi$. 
As the angle $\theta$ sets the relative orientation of the disc and the star magnetosphere we propose 
that accretion torque depends on it as 
$N_{\mathrm{acc}} \propto f(\cos\theta)$, where, according to the spherical cosine rule,
\begin{equation}
    \cos\theta = \cos\alpha\cos\chi + \sin\alpha\sin\chi\cos\psi.
    \label{eq:costheta}
\end{equation}
% and $\alpha$ is the angle between the disc normal and the NS spin axis. 
The function $f(\cos \theta)$ is unknown and depends on the details of disc-magnetosphere interaction. 
As the north and south magnetic poles are indistinguishable from the point of view of magnetohydrodynamics, the unknown function $f(\cos \theta)$ should be an even function of its argument. 

Torque modulations may be related to the changes in the configuration of the magnetosphere and its interface with the disc on dynamical time scales. 
The mean value of $N_{\rm acc}$ should not be affected by these variations, as the mean mass accretion rate is set by the boundary conditions (mass transfer rate from the donor star), and the net angular momentum depends, apart from $\dot{M}$, on the size of the magnetosphere, that is itself a weak function of inclination. 
Though in general accreting NSs should be inclined, and accretion upon inclined dipoles was studied both analytically \citep{wang_inclined, Bozzo18} and numerically \citep{romanova09}, the problem of mass accretion upon a misaligned oblique rotating dipole (where both $\alpha$ and $\chi$ are non-zero) is, to our knowledge, relatively unstudied. 
Notable exceptions are \citet{lai11} who considered the aligning torques analytically, and \citet{romanova20}, where magnetospheric accretion upon a misaligned oblique magnetic dipole is considered by means of 3D magnetohydrodynamic simulations. 
One of the interesting findings of \citet{romanova20} is that the mass accretion rate upon the simulated star varies with time at about spin period. 
% that we discuss in Section~\ref{sec:disc:warp}.
As we show below, the only case when magnetic angle changes due to accretion torque is this misaligned oblique case ($\alpha \neq 0$ and $\chi \neq 0$), and the accretion torque variations within the spin period plays crucial role in the evolution of $\chi$. 
% is the only case when magnetic angle may change due to accretion torque only. 

\begin{figure}
    \centering
    \includegraphics[width=0.8\columnwidth]{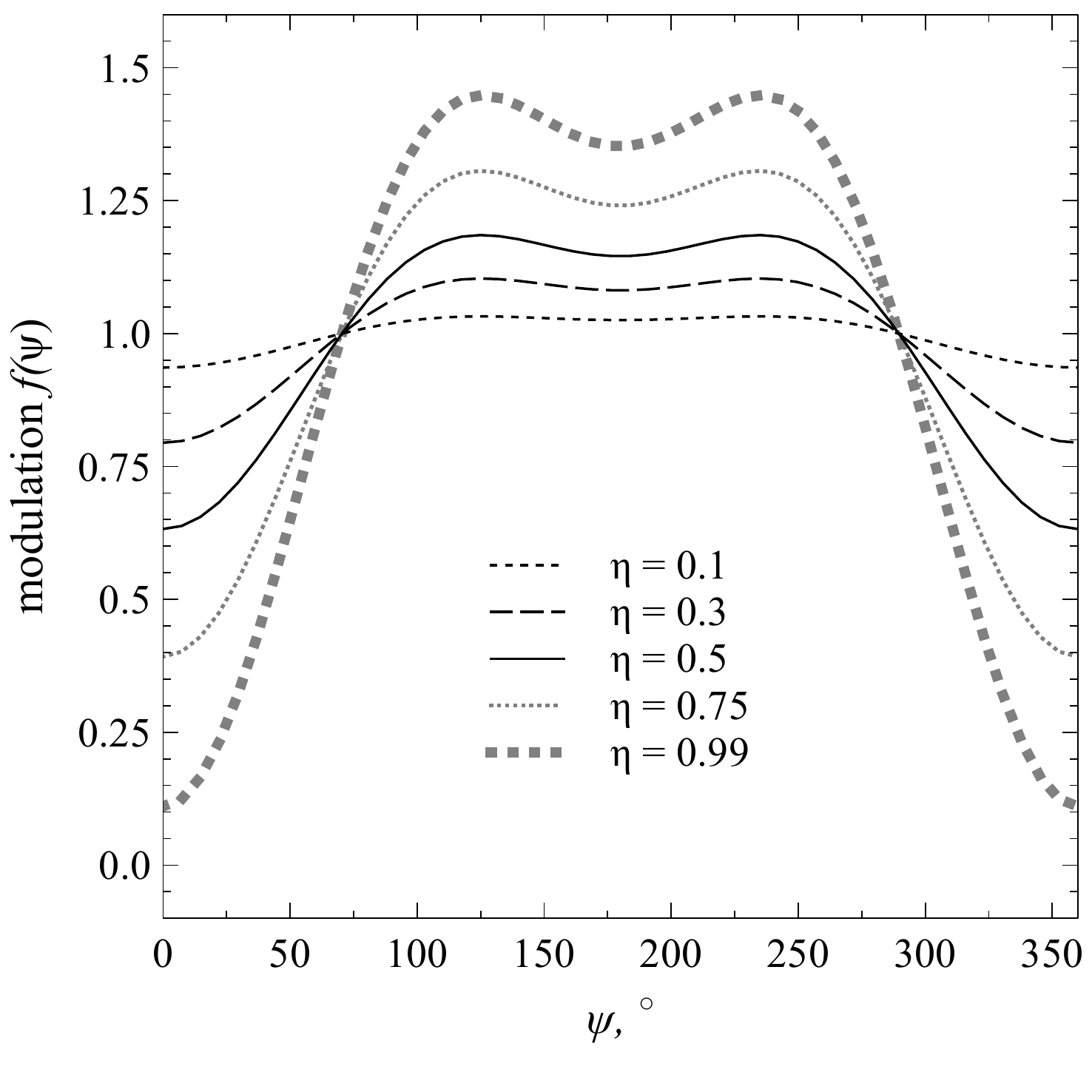}
    \caption{Accretion torque modulation factor $f(\eta, \psi)$ as a function of spin phase $\psi$ for various $\eta$. It was calculated by combining Equations (\ref{eq:mod_factor}) and (\ref{eq:costheta}). Constant $\eta$ describing the modulation amplitude was set to 0.1, 0.3, 0.5, 0.75 and 0.99. Magnetic angle $\chi$ and spin-disc angle $\alpha$ are equal to 60$^\circ$ and 45$^\circ$, respectively. 
    }
    \label{fig:mod_factor}
\end{figure}

For the modulation function $f$, we propose the simplest possible non-trivial even-order polynomial form 
\begin{equation}
    f(\cos\theta) = A(\eta, \alpha, \chi)  (1 - \eta \cos^2\theta),
    \label{eq:mod_factor}
\end{equation}
where $0 < \eta < 1$\footnote{We presume that only positive values of $\eta$ less then 1 are physically reasonable. This corresponds to the case when the instantaneous accretion torque is stronger for larger misalignment (smaller $\cos\theta$). Indeed,  it seems much easier for the matter to permeate the magnetosphere if the magnetic axis of the star lies closer to the disc plane. Moreover, $\eta <1$ means that, independently of the spin phase, accretion torque can not be directed opposite to the angular momentum in the disc.} is a constant describing the accretion torque modulation within the spin period, and $A(\eta, \alpha, \chi)$ is a normalization factor.
The latter can be calculated by averaging $f(\cos\theta)$ over the NS spin period assuming $\alpha$ and $\chi$ vary negligibly within one revolution of the star, 
\begin{equation}\label{eq:Anorm}
    A(\eta, \alpha, \chi) = \left ( \dfrac{1}{2\uppi}\int_0^{2\uppi} (1 - \eta\cos^2\theta) d\psi \right )^{-1}. 
\end{equation}
This multiplier ensures that the mean accretion torque value remains unchanged. 
Taking the integral in (\ref{eq:Anorm}), we get an analytic expression for $A$:
\begin{equation}\label{E:Acoeff}
    A(\eta, \alpha, \chi) = \left[ 1 - \dfrac{\eta}{2}(\sin^2\chi \sin^2\alpha + 2\cos^2\chi\cos^2\alpha) \right]^{-1}.
\end{equation}
The modulation factor (\ref{eq:mod_factor}) is shown as a function of spin phase $\psi$ in Figure~\ref{fig:mod_factor}. For all the calculations below we set $\eta = 0.99$ in order to estimate the maximal effect from the proposed modulation. 

Equation describing the evolution of the rotation velocity of the star may be then obtained by projecting (\ref{eq:general_eqs}) upon the instantaneous direction of $\mathbi{s}$ with subsequent averaging over the spin period
\begin{equation}
    I \langle \dot \Omega \rangle  = \langle \mathbi{N \cdot s} \rangle - \dot I \Omega,
\end{equation}
Hereafter $\langle x \rangle$ means 
\begin{equation}
    \langle x \rangle = \dfrac{1}{2\pi} \int_0^{2\pi} x  d\psi.
    \label{eq:averaging}
\end{equation}
As a result, we get, for a spherical star,
\begin{equation}
    I\left \langle \dfrac{\diff \Omega}{\diff t} \right \rangle =   N_0\cos\alpha + N_{\mathrm{mag}} + K_{\mathrm{psr}}(1 + \sin^2\chi)  - \dfrac{\diff \it I}{\diff \it t} \Omega.
    \label{eq:dot_omega}
\end{equation}
Similar equation for the evolution of the star-disc inclination angle $\alpha$ can be
obtained in the same way by projecting (\ref{eq:general_eqs}) onto the unit vector
$\mathbi{e}$ which always lies within the $\mathbi{s} - \mathbi{d}$ 
plane, orthogonal to the angular velocity vector $\mathbi{e} \perp \mathbi{s}$, and makes an acute angle with $\mathbi{d}$:
\begin{equation}
    \mathbi{e} = \left ( \begin{array}{c}
        \sin\psi \\
        -\cos\chi\cos\psi \\
        \sin\chi\cos\psi \\
    \end{array} \right ).
\end{equation}
As $\mathbi{\dot \Omega \cdot e} = -\Omega \dot \alpha$, projection along $\mathbi{e}$ results in
\begin{equation}
    I \Omega \left \langle \dfrac{d\alpha}{dt} \right \rangle = - \langle \mathbi{N \cdot e} \rangle = -N_0 \sin\alpha.
    \label{eq:dot_alpha}
\end{equation}
At last, evolution of the magnetic angle is calculated similarly by
projecting (\ref{eq:general_eqs}) onto $\mathbi{m}$.
After the averaging and using the relationship $d\Omega_z/dt = \dot\Omega\cos\chi - \Omega\sin\chi  \dot\chi$, we finally obtain
\begin{equation}
    \begin{array}{l} I\Omega \left \langle \dfrac{d\chi}{dt} \right \rangle = \eta  A(\eta, \alpha, \chi) N_0 \sin^2\alpha\cos\alpha\sin\chi\cos\chi \\
    \qquad\qquad\qquad + K_{\mathrm{ psr}}\sin\chi\cos\chi \\
    \end{array}
    \label{eq:dot_chi}
\end{equation}
%\textbf{
This equation is the main theoretical result of the paper.
%} 
The evolution of the magnetic angle, unlike that of $\Omega$ and $\alpha$, is affected by the variations of the spin-up torque within the spin period of the NS.

Note that the pulsar anomalous torque contributions to all the three equations vanish after the period-averaging procedure. 
For the equations for $\Omega$ and $\chi$, this is trivial, as pulsar torque is always orthogonal to $\mathbi{s}$ and $\mathbi{m}$. 
Multiplying the anomalous torque by $\mathbi{e}$ produces a triple product $\mathbi e \cdot (\mathbi s \times \mathbi m) = \sin\chi\sin\psi$ that becomes zero after averaging over $\psi$. 
Therefore this torque does not affect the evolution of $\alpha$ as well. 

Equations (\ref{eq:dot_omega}), (\ref{eq:dot_alpha}), and (\ref{eq:dot_chi}), together with (\ref{eq:accretion_torque}), (\ref{eq:nmag}), and (\ref{eq:pulsar_torque}) completely describe the rotational evolution of a
spherical accreting neutron star.

\subsection{Evolution of the physical parameters}\label{sec:evo:phys}

It has been shown that millisecond pulsars are on average $\sim 0.2 \Msun$ heavier than normal isolated neutron stars, that likely reflects the amount of matter accreted by the NS \citep{Zhang2011, Cheng2014}. 
This additional mass affects the inertia of the star.  

Here, we will use an approximate
phenomenological relationship between the moment of inertia of the NS and its mass, valid for a broad range of equations of state and masses \citep{bab17}
\begin{equation}\label{E:moi}
    I \simeq  10^{45}\ \left( \dfrac{M_*}{\Msun} \right) {\rm \,g\, cm^2}.
\end{equation}
The last term in (\ref{eq:dot_omega}) contributes to total spin-down as
\begin{equation}
     -\dfrac{1}{I}\dfrac{\diff \it I}{\diff \it t}\Omega \approx -3\times 10^{-14} \mbox{ } \dot M_1 M_{*,1.5}^{-1} P^{-1}_{0.01}\ \mathrm{ rad\,s^{-2}},
     \label{eq:mass_torque}
\end{equation}
where the mass of the star $M_*$ is normalised by $1.5\Msun$, and its spin period as $P_{0.01} = P/0.01$\,s. 

Another parameter affected by accretion is magnetic moment. We assume that the accreted matter changes the magnetic moment of the star as
\begin{equation}
    \mu = \mu_0 \left(1 + \dfrac{\Delta M_*}{\cal M} \right)^{-14/11}
    %\mu = \mu_0 \left( 1 + \dfrac{\Delta M_*}{5.64\times 10^{-3} %M_{\sun}}\right)^{-14/11},
    \label{eq:mp}
\end{equation}
where $\mu_0$ is the initial magnetic moment, $\Delta M_*$ is the amount
of the matter accreted by the star and
\begin{equation}
   {\cal M} = 1.1\times 10^{-5}  \dot M_1^{1/7}  \mu_{0,30}^{3/14} R_\mathrm{12.5}^{3}\,\Msun.
\end{equation}
Here, initial magnetic moment $\mu_{0,30}$ is expressed in the units of $10^{30}$\,G\,cm$^3$, and the radius is normalised as $R_\mathrm{12.5} = R_*/12.5$ km. 
The law (\ref{eq:mp}) approximately reproduces the magnetic field burial scaling predicted by \citet{MP01} for Eddington mass accretion rate and pulsar-scale magnetic field (the mass scale depends on mass accretion rate, as well as on magnetic field, very weakly). 

At the same time, one may expect stellar magnetic field decay due to dissipative processes within it crust \citep{reis03}. 
However, we have not taken this effect into account, as the field burial time scale 
\begin{equation}
    \tau_{\mathrm{bur}} = - \dfrac{\mu}{\dot \mu} \approx 3.4 \times 10^3  \dot M_1^{-6/7}  \mu_{0,30}^{3/14}  R_\mathrm{12.5}^{3}\ \yr
\end{equation}
is significantly shorter than the decay time scale $\tau_{\mathrm{dec}} \gtrsim 10^6$yr \citep{gullon14, ip15, bakb17}. 

%\textbf{
There are other mechanisms possibly important for the rotational evolution of NSs, that we do not consider in this paper.
As we already mentioned in Section~\ref{sec:genmodel}, non-sphericity of the star is probably a minor issue, though deformation in the direction non-coplanar with $\mathbf{s}$ and $\mathbf{m}$ may lead to free precession and a more complicated evolutionary path. 
Presence of higher multipoles, especially misaligned with the dipole, is another important factor that requires a separate study.
The importance of higher multipoles was stressed, for example, by \citet{Petri2019}. 
Probably, the most relevant contribution of higher multipoles would be the deformation of the NS not aligned with the axis of the magnetic dipole. 
In our calculations, we consider NS as a rigid body, that may be incorrect if the core of the star is weakly coupled to the crust \citep{Casini1998}. 
If the crust itself is non-rigid (as in the model by \citealt{Ruderman1991}) or the magnetic field lines are allowed to slip through the crust, the rotational evolution of the NS should also become much more complicated.  
All these effects, however, are not usually included in the calculations of NS evolution due to the large uncertainties in their parameters. 
%}
\begin{comment}
And, in principle, there are other certain purely studied effects that possibly
important for the rotational and magnetic evolution of a neutron star. We have already
mentioned the non-sphericity of the star in the Section~\ref{sec:genmodel} which is
likely to be weak enough to be avoided for our purposes. On the other hand, multipolar magnetic field components \citep{Petri2019}, NS core-crust coupling and effects of the strong gravity \citep{Casini1998} or even ``tectonics'' of the star surface \citep{Ruderman1991} may also eventually affect upon the evolution of the magnetic angle.
All these phenomena, however, are not usually included to the classical consideration of NS evolution due to large uncertainty of their details. Thus we also avoid them in the framework of our calculations. In particular, we restricted ourselves with a rigid-body approach and consider neutron star with dipole magnetic field.
\end{comment}
%______________________________________________ 

\section{Results}
\label{sec:res}

\begin{table}
\caption{Parameters of the NS evolution models considered in the paper. Initial spin period $P_0 = 2\pi/\Omega_0$. Modulation constant $\eta = 0.99$ are the same for all models. Initial magnetic moment $\mu_{0,30}$ is normalised on $10^{30}$\,G\,cm$^3$, while accretion rate $\dot{M}_1$ is taken in Eddington rates $\dot{M}_{\rm Edd, {\sun}}$ for one Solar mass.}
\label{tab:models}  
\centering          
%\begin{tabular}{r r r r r r}
\begin{tabular}{r r r r}
\hline        

Model & $P_{0}$, sec & $\mu_{0,30}$ & $\dot M_1$ \\  % table heading 
\hline                        % inserts single horizontal line
    A & 0.1 & 5.0 & 0.01 \\
    B & 0.1 & 5.0 & 1.0 \\
    C & 5.0 & 1.0 & 1.0\\
    D & 100.0 & 1.0 & 1.0 \\
\hline                                   %inserts single line

\end{tabular}
\end{table}

Bringing together all the physics described in the previous section allows to reconstruct the rotational evolution of an accreting NS, including magnetic angle (Equation~\ref{eq:dot_chi}). In our model, there are two processes affecting $\chi$: the modulation term and pulsar losses. While the last is well known in the context of the evolution of radiopulsars, it is usually not important during the accretion stage. The modulation term, however, is proportional to $\sin^2\alpha$ and thus is only important during the rotation axis alignment stage. 
The resulting change in $\chi$ may be under favourable circumstances comparable to the initial misalignment $\alpha$ and important for the resulting distribution in magnetic angles. 

In this Section, we consider four different scenarios of NS evolution in binary systems undergoing mass transfer at a fixed rate. 
The models differ in the initial conditions and in the mass transfer rates. 
Their parameters are given in Table~\ref{tab:models}. 
The initial mass of the NS in all the models is $1.5\Msun$. Modulation factor $\eta$ for the accretion torque was everywhere fixed to 0.99. 
We use the radius of the NS equal to 12.5km, but this does not affect the results much as the radius of the magnetosphere is always larger. 
The moment of inertia is given by Equation~(\ref{E:moi}), and the magnetic moment of the star decreases with time according to the burial law~(\ref{eq:mp}).

We track the evolution for $10^7-10^8$\,yr depending on the model until the mass of the NS changes considerably (up to $\sim 0.1\Msun$). 
The four models differ in the magnetic moment, mass accretion rate, and initial NS spin, and may be loosely associated with the different scenarios of NS rotational evolution. Lower mass accretion rates are expected in binary systems with lower-mass donor stars \citep{AMSP21}.

Note that we do not pretend here on capturing all the details of the binary evolution. 
In combination with binary evolution codes, out results will hopefully allow to reproduce a realistic picture of the evolving magnetic angle distributions in real X-ray binary populations. 

First model (A) corresponds to the case of a putative MSP progenitor with a low-mass donor. 
Indeed, accreting millisecond pulsars usually have very low mass donor stars and relatively low mass accretion rates $\sim 0.01-0.1$ $\dot{M}_{\rm Edd,\, {\sun}}$ \citep{AMSP21, tauris2012}. 
At the same time, accretion at a mean rate smaller than $\sim 0.01M_{\rm Edd,\, {\sun}}$ is insufficient to spin up an average MSP progenitor to a typical observed period of $\sim 10$ms \citep{pan2013}.

In this model initial parameters of the NS are similar to those of a newly born radiopulsar: spin period $\sim 0.1$s and initial surface magnetic field $5\times 10^{12}$G. The adopted mass accretion rate is $0.01 \dot M_{\mathrm{Edd},{\sun}}$, that allows to track the rotational evolution for more that tens of millions of years. 

The second model (B) describes a similar situation but with a larger mass accretion rate of about Eddington. This is closer to the evolution of an X-ray pulsar, though real X-ray pulsars also show strong modulations (orbital and stochastic) of mass accretion rate we did not include in the model. We consider chosen $\dot M$ as an average accretion rate within the system evolutionary timescale. 

The remaining two models (C and D) explore the case of an initially slowly rotating NS spun up at the Eddington rate. 
Thus, model D may represent the case of the rapid spin-up observed for NGC 300 ULX1 \citep{carpano18, vasilopoulos}.

For each of the models, we have considered different initial values of $\alpha$ and $\chi$, arranged in a two-dimensional grid with a step of one degree along each coordinate. 

\begin{figure*}
    \centering
    \includegraphics[width=\textwidth]{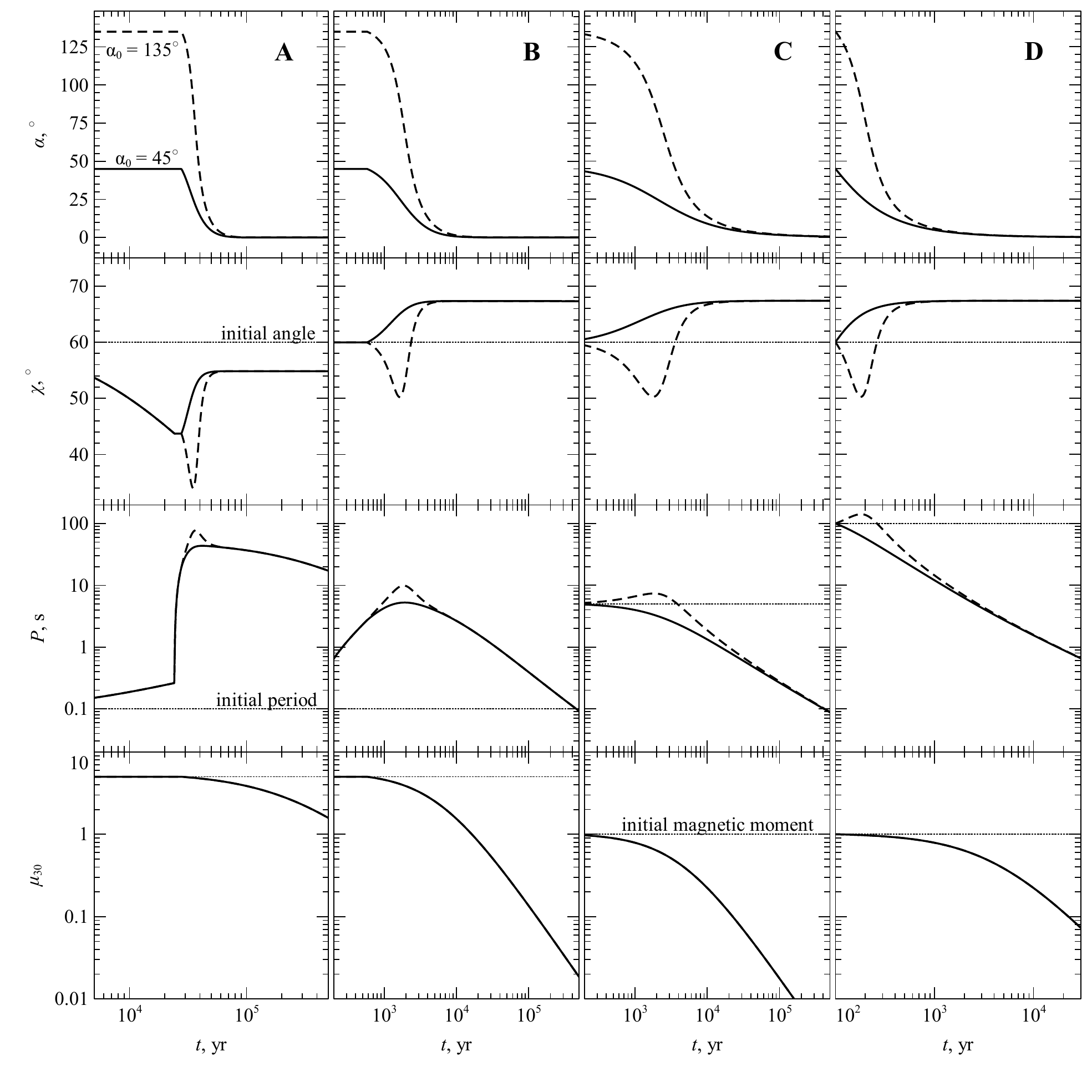}
    \caption{Rotational evolution of accreting NSs, according to the models A-D listed in Table~\ref{tab:models}. 
    Solid lines represent the cases with initial $\alpha_0 = 45^{\circ}$, dashed are for $\alpha_0 = 135^{\circ}$. 
    Thin dotted lines mark the initial values of the quantities. 
    % Modulation constant $\eta$ was set to 0.99 for all these models. 
    Dipole magnetic moment of the star $\mu_{30}$ is normalised to $10^{30}{\rm \,G\, cm}^3$.} 
    \label{fig:models_evolution}
\end{figure*}

In Figure~\ref{fig:models_evolution}, selected evolutionary trajectories are shown.
For each model, we show the cases with initial magnetic angle $\chi_0 = 60\degr$ and two inclination angles, $\alpha_0 = 45\degr$ and $\alpha_0 = 135\degr$. 
For all the models, a short-term ($\sim 10^4$ years for model A, and about $10^3$ \yr\ for all the others) magnetic angle counter-alignment episode is clearly seen at the beginning of the accretion stage. 

The evolution of $\chi$ is related either to this spin-disc alignment stage or to the pulsar torque.
For the model A, the NS starts as an ejector, and the rotational evolution for the first $\sim 25$ kyr is indistinguishable from that of a classic radiopulsar. 
Model B starts with a propeller stage several hundred years long. 
On average, $\chi$ increases by several degrees (up to $\Delta\chi \sim 15\degr$) during the alignment stage and settles at a constant level when $\alpha$ stops evolving. 

Indeed, the second term in (\ref{eq:dot_chi}) vanishes as $\alpha \to 0$, while the pulsar torque is inefficient due to the small spin period and buried magnetic field. 
The contribution of the pulsar torque may however become important if the torque is enhanced by the opening of magnetic field lines considered by \citet{parfrey16}. 
We discuss this possibility in Section~\ref{sec:disc:pulsar}.

If the NS initially rotates backwards with respect to the angular momentum of the disc ($\alpha_0 > 90\degr$), the coupled evolution of $\chi$ starts with magnetic alignment, that is then (as $\alpha$ becomes smaller) compensated by rapid orthogonalisation, resulting in identical final magnetic angle values for $\alpha_0$ and $90\degr-\alpha_0$. 
As we will show in Section~\ref{sec:disc:analytic}, this is a general result valid as long as we neglect pulsar losses.

\begin{figure*}
    \centering
    \includegraphics[width=\textwidth]{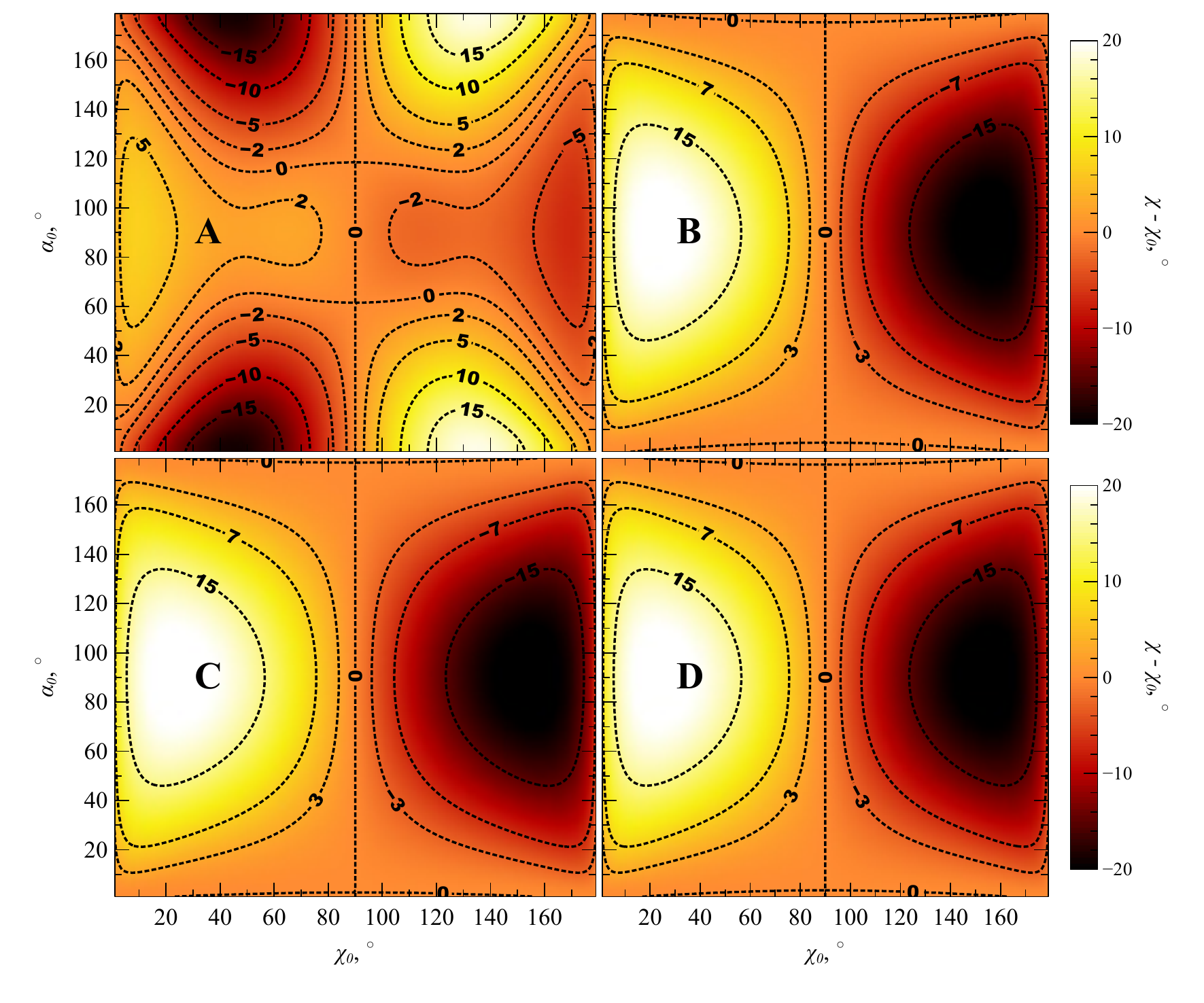}
    \caption{Difference in magnetic angles  $\Delta \chi = \chi - \chi_0$ after $10^5$ years of evolution for different models as function of $\chi_0$ and $\alpha_0$. 
    Top left plot is for model A, top right -- B, bottom left -- C and bottom right -- D respectively (see Table~\ref{tab:models} for details.) }
    \label{fig:models_net}
\end{figure*}

In Figure~\ref{fig:models_net}, we show the overall change in magnetic angle $\Delta \chi = \chi - \chi_0$ after $10^5$ years of evolution for different $\chi_0$ and $\alpha_0$. 
It is clearly seen that for all the models, there is systematic magnetic orthogonalisation. 
Under our assumption of relatively strong accretion torque modulation, magnetic angle $\chi$ changes for up to $15$ degrees during the spin-disc alignment stage and then remains constant unless pulsar losses become important.

The mirror symmetry of the plots relative to the $\chi_0 = 90\degr$ line is related to the symmetry of the magnetic poles. Both positive values of $\Delta \chi$ at $\chi_0 < 90\degr$ and negative at $\chi_0 > 90\degr$ correspond to the increasing misalignment of the magnetic axis with the axis of rotation. 

\begin{figure*}
    \centering
    \includegraphics[width=\textwidth]{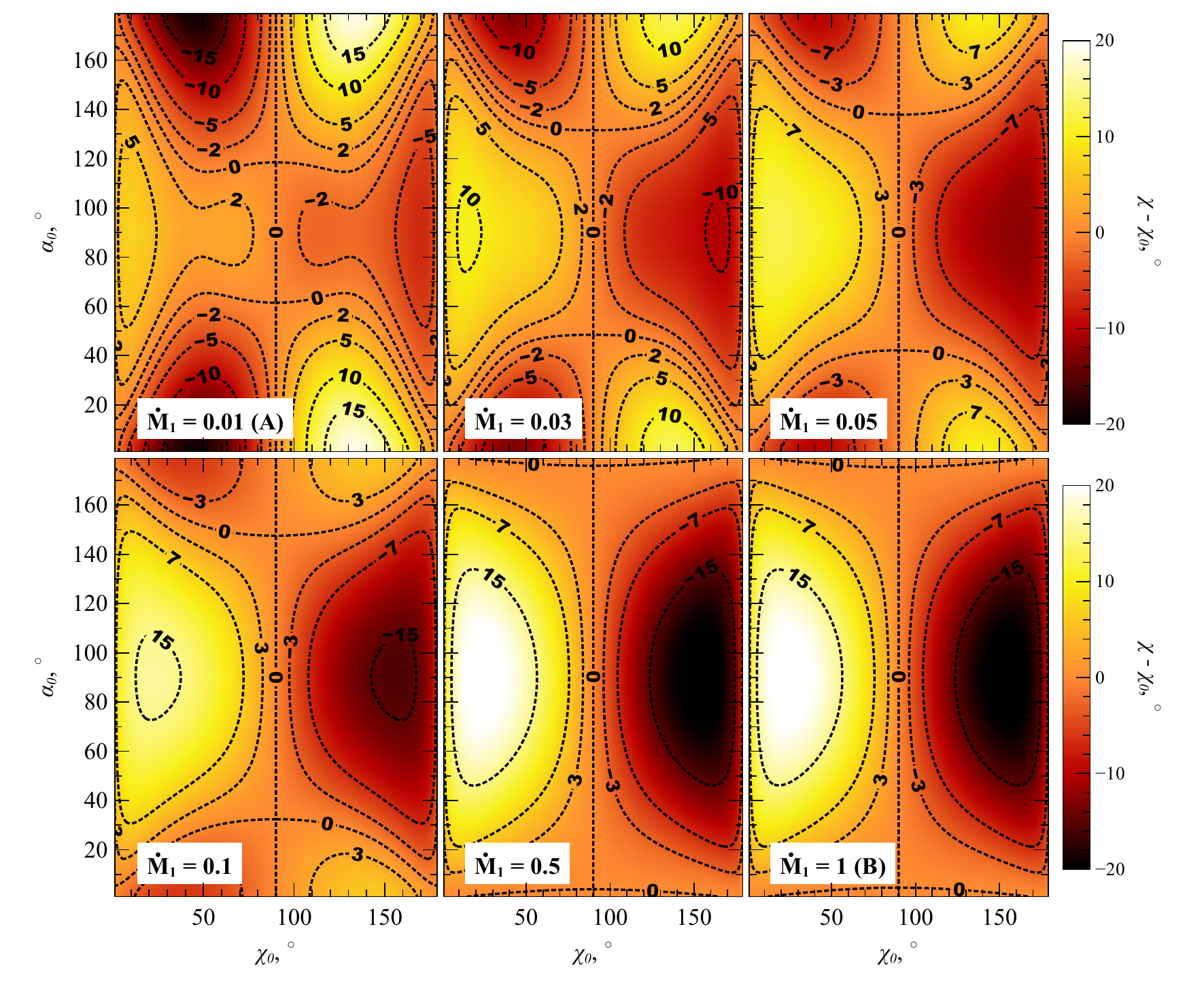}
    \caption{The same as  Figure~\ref{fig:models_net} for models with the mass accretion rates intermediate between models A and B. All the other parameters are the same as in A/B. 
    }
    \label{fig:accretion_net}
\end{figure*}

%\textbf{
In addition, in Figure~\ref{fig:accretion_net} we show how the
difference between final and initial magnetic angles depends upon the adopted accretion rate $\dot M_1$. 
Namely, we calculated $\Delta \chi$ with the initial parameters of model A, but for different values of $\dot M_1$ = 0.01, 0.03, 0.1, 0.5, and 1.0. 
First of these values corresponds to the original model A as it is defined in Table~\ref{tab:models}, while the last one is identical to model B. 
The maps shown in Figure~\ref{fig:accretion_net} outline the result of the first $10^5$ years of evolution. 
The map changes smoothly with the increasing mass accretion rate, mostly due to decreasing contribution from pulsar losses, that in our model tend to compensate the orthogonalisation during the spin alignment stage.
The larger the mass accretion rate, the smaller is the overall contribution of the pulsar losses. 
The last two panels of the figure are practically unaffected by pulsar losses, and show the effect of the coupled evolution during the alignment stage.
%}

\section{Discussion}
\label{sec:disc}

\subsection{Pulsar losses during accretion}
\label{sec:disc:pulsar}

Should pulsar torques remain the same during the accretion stage? 
On one hand, accretion is often claimed to interfere with pulsar emission mechanisms \citep[e.g.][]{Shvartsman1971, Melatos2016_Poison, Papitto2020_TPSR}. 
On the other hand, independently of the physical conditions in the magnetosphere, open magnetic field lines should still provide a torque determined by the magnetic stresses at the light cylinder \citep{cs06}. 

As it was shown by \citet{parfrey16} for the case of an aligned rotator, the presence of a conducting disc opens up some of the closed magnetic field lines and thus increases the spin-down by a factor proportional to the squared magnetic flux through the open field lines. 
For an accreting NS with $r_{\rm m} \sim r_{\rm co} \ll r_{\rm LC}$, this results in $N_{\rm Parf} \sim N_{\rm psr} \left( r_{\rm LC}/ r_{\rm m}\right)^2 \sim N_{\rm psr} \left( r_{\rm LC}/ r_{\rm co}\right)^2 \left(r_{\rm co} / r_{\rm m} \right)^2 \sim N_{\rm mag} \left( r_{\rm co}/r_{\rm LC}\right) \left(r_{\rm co} / r_{\rm m} \right)^2$. The contribution of this enhanced spin-down torque is normally smaller than the estimated magnetospheric losses unless the NS is far from equilibrium rotation and $r_{\rm m} \ll r_{\rm co}$. 
But, if the aligning pulsar torque (the second term in Equation~\ref{eq:pulsar_torque}) is scaled similarly, magnetic angle very rapidly evolves to zero during the accreting millisecond pulsar stage.

\begin{figure}
    \centering
    \includegraphics[width=1\columnwidth]{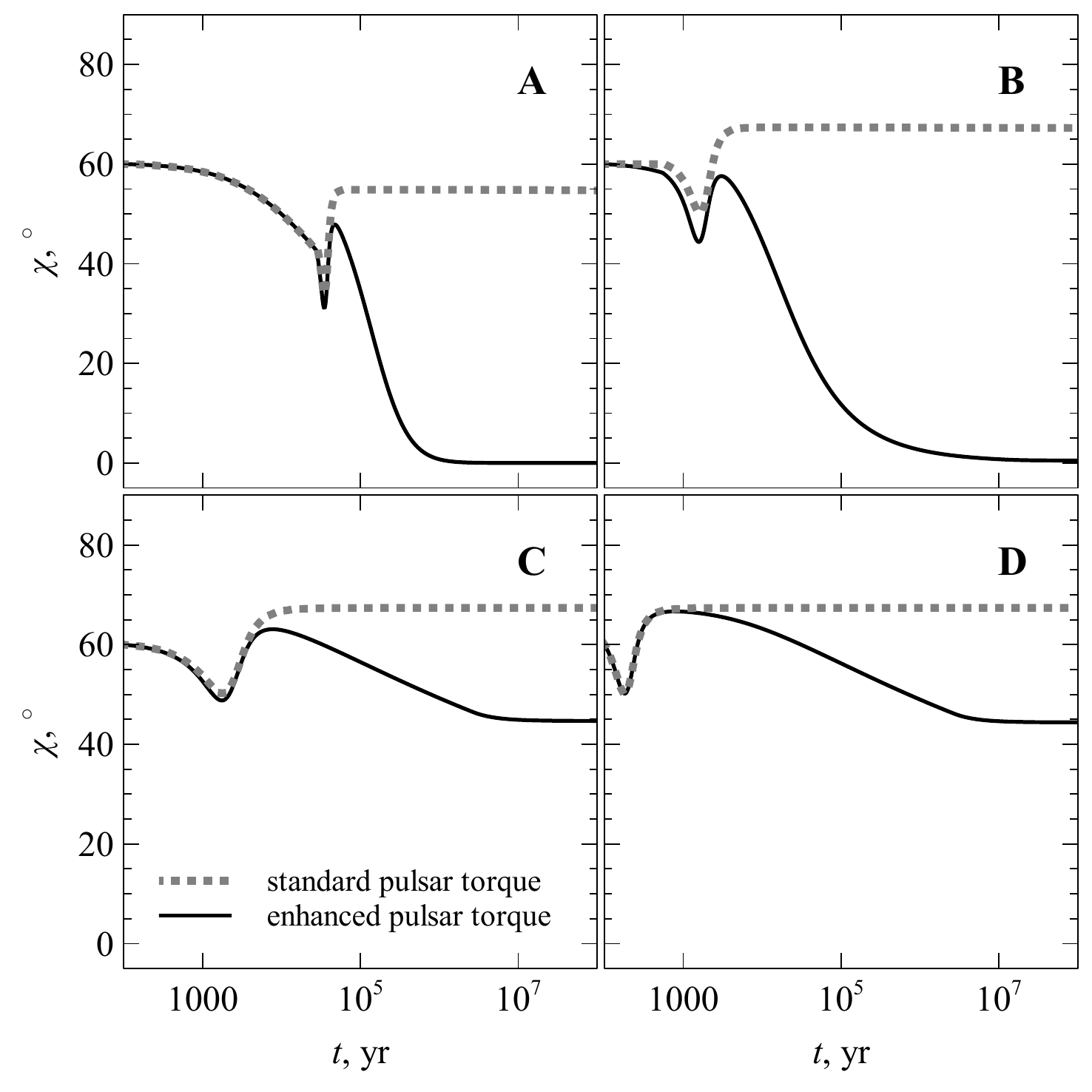}
    \caption{Magnetic angle evolution when magnetic alignment torque enhancement described in \citep{parfrey16} is taken into account. For each considered model from the Table~\ref{tab:models}, the case of $(\chi_0, \alpha_0)$ = $(60^\circ, 135^\circ)$ is shown. 
    The evolution without enhancement is shown by grey dashed lines, while enhancement scenarios are represented by solid black lines.}
    \label{fig:parfrey}
\end{figure}

Indeed, as pulsar obliquity evolution is driven physically by the same torque as the pulsar spin-down, one would expect the magnetic alignment time scale $\tau_\mathrm{psr}$ (\ref{eq:time_psr})
to decrease by the same factor as the spin-down time scale.

So, torque enhancement decreases it by the factor
\begin{equation}
    \left( \dfrac{r_\mathrm{m}}{r_\mathrm{LC}} \right)^2 \approx 
    3.5\times 10^{-3}  \mu_{30}^{8/7}  {\dot M_1}^{-4/7}  \left ( \dfrac{M_*}{\Msun}\right)^{-2/7}  P_{\rm s}^{-2}.
    \label{eq:time_psr_enh}
\end{equation}
Taking this effect into account leads to faster evolution of $\chi$ in an accreting system. 
In Figure~\ref{fig:parfrey} we show the evolution of $\chi$ for all the models from Table~\ref{tab:models}. 
Initial conditions are $\chi_0 = 60^{\circ}$ and $\alpha_0 = 135^{\circ}$ that allows comparison with Figure~\ref{fig:models_evolution}.
Quite expectedly, the impact of the enhancement factor strongly depends on the initial rotation period and on the magnetic field of the star. 
If the initial spin period of the NS is relatively large ($P_0 \gtrsim 1\,$s, models C and D), then the star is additionally aligned by $\sim 20\degr$ during the accretion stage and then stops due to magnetic field burial. On the other hand, for the relatively small initial period $P_0 = 0.1$\,s (models A and B), magnetic angle decreases effectively to zero before the expected magnetic field burial time. 

We conclude that magnetic alignment torque enhancement, if present, is a very strong effect. 
If present on the spin-up stage in MSP progenitors, it should probably make all the magnetic angles in these sources too small for the MSPs to be observed as pulsars. 
However, more detailed analysis of this issue is required.

\subsection{Analytic solution for the coupled orthogonalisation process}\label{sec:disc:analytic}

In our approach, there are two main processes affecting magnetic angle: magnetic alignment by pulsar torques and $\chi-\alpha$ coupling during the alignment of the rotation axis with the disc. 
First effect is well studied and plays important role in the evolution of normal
radiopulsars on their spin-down time scales (typically $\sim 10^6-10^8$ years). 
For accreting systems, the most important stage of $\chi$ evolution is the alignment of the rotation axis. 
The overall change of $\chi$ may be comparable to $\alpha$ but strongly depends on the modulation within the spin phase. 
If we neglect the pulsar torque and exclude time from equations~(\ref{eq:dot_alpha}) and ~(\ref{eq:dot_chi}), magnetic angle becomes effectively a function of $\alpha$ only
\begin{equation}\label{E:res:dchidalpha}
    \frac{d\chi}{d\alpha} = -\eta A(\eta, \alpha, \chi) \sin\alpha \cos\alpha \sin\chi \cos\chi, 
\end{equation}
where $A(\eta, \alpha, \chi)$ is given by Equation~(\ref{E:Acoeff}).
Note that, as $A$ does not change with the changing sign of $\cos \alpha$, and the differential of $\diff \alpha$ appears in combination $\cos\alpha \sin\alpha \diff \alpha$, any solution of Equation~(\ref{E:res:dchidalpha}) will have the symmetry between $\alpha$ and $\uppi - \alpha$. 
After the end of the alignment stage, the solution $\chi(\alpha = 0)$ depends only on $\chi_0$ and $\sin\alpha_0$, that justifies the independence of the final results on the sign of $\cos \alpha_0$ in Figure~\ref{fig:models_evolution}. 

For $\eta \ll 1$, Equation~(\ref{E:res:dchidalpha}) may be integrated analytically, providing an expression for the evolution of $\chi$
\begin{equation}
\displaystyle     \tan \chi \simeq \tan \chi_0 \exp\left\{ \frac{\eta}{2} \left(\cos^2\alpha - \cos^2\alpha_0\right)\right\}.
\end{equation}
Or, in terms of $\Delta \chi = \chi - \chi_0$
\begin{equation}
    \Delta \chi \simeq \dfrac{\eta}{2}\cos^2\chi_0 \left( \cos^2\alpha -\cos^2\alpha_0\right). 
\end{equation}
The changes in magnetic angle depend only on initial inclination, initial magnetic angle, and spin-phase modulation. 
As cosine is a decreasing function, $\chi$ increases with alignment if and only if $\alpha < 90\degr$. 

\subsection{Spin-disc alignment from the observational point of view}

The evolution of a NS in a binary system may consist of one or more $\chi$ alignment episodes driven by pulsar torque, and one or more magnetic angle leaps due to modulation coupling with $\alpha$.
The time scales for these stages are profoundly different. Pulsar magnetic alignment time is approximately the same as the pulsar spin-down time (\ref{eq:time_psr}), while rotation axis is aligned with the disc on the spin-up time scale (\ref{eq:time_align}).

Normally, spin-disc alignment of the NS is a very short stage compared to the lifetime of the source. 
Besides, the overall change in magnetic angle is limited by a quantity of about $\Delta \chi_{\rm max} \lesssim \frac{\eta}{2}\Delta \alpha$. 
In favourable conditions, if the modulation factor $\eta \simeq 1$ and initial $\alpha_0 \sim 20\div 40$ or $140\div 160\degr$, we expect the magnetic angle to change by a considerable amount of 10$\div$15 degrees (see Figure~\ref{fig:models_net}).

Apparently, such alignment episodes sometimes happen. An important example is NGC300~X-1 \citep{carpano18}, known as a peculiar Be/X-ray binary with a NS spun up from about 100 to 20s in several years in an episode of rapid accretion \citep{vasilopoulos}. 
Such a strong change in observed period allows for an equally dramatic change in the direction of the axis of rotation. 
In fact, the state in which the system existed before the outburst of 2010 suggests that the NS was fed by a low-density wind carrying little angular momentum.

\subsection{Wind accretion}

During wind accretion, the direction of the angular momentum captured by the NS is not as stable as in the case of accretion through the inner Lagrangian point. 
As it was proposed by \citet{wang81} and qualitatively confirmed by numerical simulations \citep{FT88, ruffert97, ruffert99,MMRR}, inhomogeneities in density and velocity of the wind make the direction of the angular momentum in the disc or in the quasi-spherical accretion flow (as in \citealt{shakura12}) a random quantity changing on the time scales much smaller than the alignment time scale. 
In this case, the equilibrium rotation period of the NS should be longer than for disc-fed accretion at equal rate, and there is no real alignment of the rotation axis. 
Instead, each separate episode of stable spin-up is associated with evolution (alignment) of $\alpha$ and related modulation-mediated evolution of $\chi$. 

On the time scales smaller than the time scale of the variations in the inflowing angular momentum, Equation~(\ref{E:res:dchidalpha}) is still valid. 
On longer time scales, the evolution of magnetic angle is described by a stochastic equation following from (\ref{eq:dot_chi}) in the absence of pulsar losses
\begin{equation}\label{E:disc:sto}
    \frac{d}{dt}\left(\ln \tan \chi\right) = \eta A \frac{N_0}{I\Omega} \sin^2\alpha \cos\alpha
\end{equation}
where $\alpha$ is now a random function of time. 
We can think of the evolution of $\ln \tan\chi$ as a random walk process caused by a white noise in the right-hand side of Equation~(\ref{E:disc:sto}). 
Let us consider that the direction of the external angular momentum $\mathbi{N}$ is stable on some time scale $\tau_\alpha \ll \tau_{\rm align}$, and completely random and isotropic on longer times.

If the orientation of the inflowing angular momentum is random and isotropic, the mean value of the right-hand side of Equation~(\ref{E:disc:sto}) is zero, and there is no average trend in magnetic angle evolution. 
However, dispersion of the derivative of $\ln \tan \chi$ is non-zero and independent of $\chi$, that means dispersion of the quantity itself grows with time approximately linearly, as
\begin{equation}\label{E:disc:disp}
\begin{array}{lcc}
 \displaystyle      \left\langle \left(\ln \tan \chi\right)^2\right\rangle 
    &\simeq &  \displaystyle    \eta^2 A^2 \left(\frac{N_0}{I\Omega}\right)^2 \left\langle \left(\cos^2\alpha \sin\alpha \right)^2\right\rangle \tau_\alpha t \\
 \displaystyle    & = &  \displaystyle    \frac{8}{105} \eta^2 A^2 \left(\frac{N_0}{I\Omega}\right)^2 \tau_\alpha t,\\
\end{array}
\end{equation}
where $\tau_\alpha$ is the characteristic correlation time scale of $\alpha$.
Magnetic angle is involved in a random walk process that is linear in $\ln \tan \chi$, and thus tends to form a uniform distribution in this quantity. Uniform distribution in $\ln \tan \chi$ corresponds to 
\begin{equation}
    \displaystyle \frac{dN}{d\chi} \propto \frac{1}{\left|\sin\chi\cos\chi\right|}.
\end{equation}

We expect such a stochastic regime to be realised in HMXB systems where wind-fed accretion is the most probable scenario at low mass accretion rates (see for instance \citealt{chaty11}). 
Sufficiently long accretion in this regime will leave the NS with a random orientation of the rotation axis, but the distribution in magnetic angle will have much more objects with $\chi$ close to either zero (or 180\degr, as the magnetic poles are indistinguishable) or 90\degr. 
Thus, we expect many wind-accreting HMXBs to have nearly-aligned or nearly-orthogonal NSs. 

\subsection{Magnetic orthogonalisation due to magnetosphere-disc interaction}\label{sec:disc:warp}

\cite{lai11} have considered interaction of a protoplanetary disc with a magnetised star. Such a system is physically similar to that considered in this work. Unlike most of the studies in the field, \cite{lai11} assumed that both the magnetic moment and the angular momentum in the disc are inclined with respect to the spin axis of the star.

Though the initial assumptions of this paper are based on a simple resistive model similar to \citet{GL77}, their results give the impression of what may be the effect of disc interaction with the magnetosphere. 
The real penetration depth is likely much smaller than in resistive models, implying that the effect should also be not so profound.

According to \citet{lai11}, the spin of the central star is affected by a 
back-reaction torque owing to the interaction of the surface currents in the disc with the poloidal field of the star. 
The torque may be split into two components, warping and precession.
Warping torque equals
\begin{equation}
    \mathbi{N}_\mathrm{warp} = \dfrac{1}{6} \xi^{-7/2} N_0 \cos^2\chi \cos\alpha \ (\mathbi{d} \times (\mathbi{s} \times \mathbi{d})).
\end{equation}
After averaging over the spin period, it introduces additional terms to Eqs~(\ref{eq:dot_omega}) and (\ref{eq:dot_alpha}),
\begin{equation}
    I\left \langle \dfrac{d
    \Omega}{dt} \right \rangle_{\mathrm{warp}}= \dfrac{N_0}{6} \xi^{-7/2} \cos^2 \chi \sin^2 \alpha \cos \alpha
    \label{eq:dot_omega_warp}
\end{equation}
and
\begin{equation}
     I \Omega \left \langle \dfrac{d
    \alpha}{dt} \right \rangle_{\mathrm{warp}} =   \dfrac{N_0}{6} \xi^{-7/2}\cos^2 \chi \cos^2\alpha \sin\alpha,
    \label{eq:dot_alpha_warp}
\end{equation}
but does not contribute to Equation~(\ref{eq:dot_chi}) describing the evolution of magnetic angle.
The other, precessional, back-reaction torque vanishes after averaging for the same reasons as the other precession torques (see Section~\ref{sec:modulation}).

The aligning warping torque given by Equation~(\ref{eq:dot_alpha_warp}) is positive for $\alpha < \uppi/2$ and thus slows down the alignment of the spin axis. 
 
\begin{figure*}
    \centering
    \includegraphics[width=2\columnwidth]{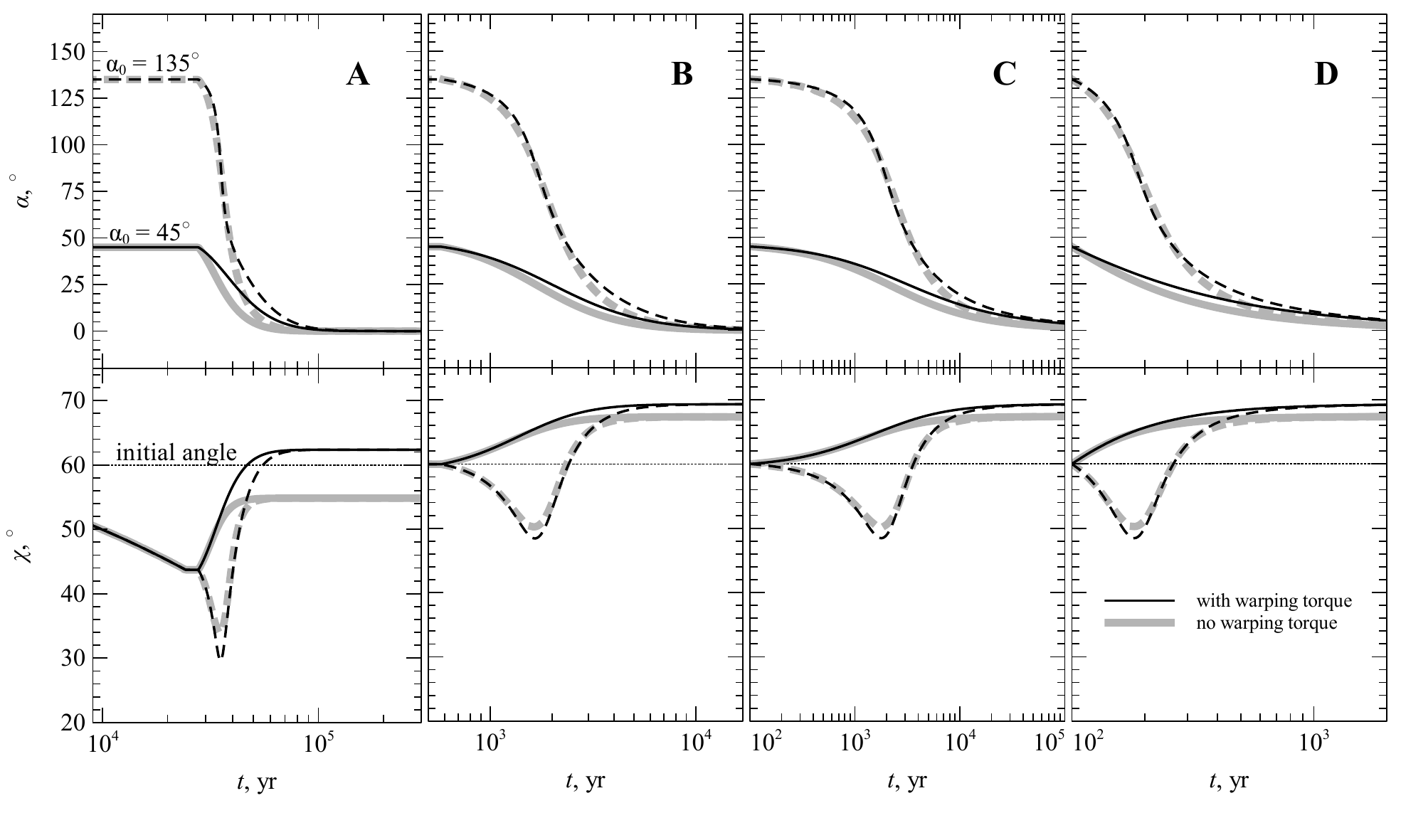}
    \caption{Magnetic angle evolution taking into account back-reaction 
    warp torque described in \citep{lai11}. We show the same set of models as in Figure~\ref{fig:models_evolution}. Alignment history without warp torque is shown by thick grey lines in each panel, while warp scenarios are represented by black lines. Dashed lines are used for $\alpha_0=135\degr$, solid for $\alpha_0 =45\degr$.}
    \label{fig:warp}
\end{figure*}

We have included the warping terms (\ref{eq:dot_omega_warp}) and (\ref{eq:dot_alpha_warp}) to our simulations and additionally run all the four models with them. 
The resultant NS spin evolution is shown in Figure~\ref{fig:warp}. 
It is clearly seen that adding a warping torque allows $\alpha$ to remain relatively large for a longer time. 
As the magnetic angle evolves due to the modulation term at a rate proportional to $\sin^2\alpha \cos\alpha$ (see Equation~\ref{eq:dot_chi}), magnetic angle orthogonalisation becomes more profound.
Depending on the particular initial configuration, $\Delta \chi$ may reach values  several times larger than during conventional evolution considered in our paper. 
Maximal values of $|\Delta\chi| \simeq 50\degr$ are observed for initially small magnetic angles $\chi_0 \lesssim 5-10^\circ$ for a broad range of $\alpha_0$.

However, the contribution of the warping torque should be treated with caution, as it relies on the rather poorly known details of magnetic field interaction with the disc. 
In particular, it treats the disc as planar, while a real accretion disc around an inclined rotator, being subject to warping torques, should become warped.
A more elaborate version should take into account the shape of the disc as well as for a more realistic dynamics of disc-magnetosphere interaction, involving instabilities and magnetic field reconnection.

%
%______________________________________________________________

\section{Conclusions}\label{sec:conc}

The main issue we addressed in this paper was whether it is possible to change the magnetic angle of a NS by accretion. 
We find this possible under quite specific circumstances: when the rotation axis is misaligned with the spin-up torque, and the spin-up torque itself is a function of the spin phase. Under favourable conditions, spin-disc alignment leads to misalignment of the magnetic axis, but always at a lower rate. 

If the direction of the angular momentum of the accreting matter is stable (as in the case of Roche-lobe overflow), this alignment is only a short episode in the beginning of accretion. 
Outside this short alignment stage, evolution of the magnetic angle is affected only by pulsar aligning torque or its modifications. 
It is unclear if the aligning torque should be enhanced due to the opening of magnetic field lines by the accretion disc. 

Accretion from a strongly inhomogeneous stellar wind results in a different picture. 
Strong variations of velocity and density make the net angular momentum of the accreting matter a random quantity. 
We show that, in this case, magnetic angle experiences a random walk, related to the random-walk behaviour of the rotation axis. 
The rate of this random walk depends strongly on the magnetic angle itself, leading to a higher probability to find an object in either aligned or orthogonal state. 

In all the scenarios, the rate of magnetic angle alignment is proportional to $\sin\chi\cos\chi$, suggesting that any distribution in $\chi$ sufficiently affected by accretion should have an excess of aligned and orthogonal rotators. 
However, the full picture is more complex and requires taking into account the initial distribution in magnetic angle, accretion histories, and the correct form of the spin-down torque during the accretion stage. 

\section*{Acknowledgements}

AB acknowledges the Russian Government Program of Competitive Growth of Kazan Federal University supporting the work. 

\section*{Data availability}
The data underlying this article will be shared on reasonable request to the corresponding author.

\bibliographystyle{mnras}
\bibliography{eman}

\begin{thebibliography}{}
\makeatletter
\relax
\def\mn@urlcharsother{\let\do\@makeother \do\$\do\&\do\#\do\^\do\_\do\%\do\~}
\def\mn@doi{\begingroup\mn@urlcharsother \@ifnextchar [ {\mn@doi@}
  {\mn@doi@[]}}
\def\mn@doi@[#1]#2{\def\@tempa{#1}\ifx\@tempa\@empty \href
  {http://dx.doi.org/#2} {doi:#2}\else \href {http://dx.doi.org/#2} {#1}\fi
  \endgroup}
\def\mn@eprint#1#2{\mn@eprint@#1:#2::\@nil}
\def\mn@eprint@arXiv#1{\href {http://arxiv.org/abs/#1} {{\tt arXiv:#1}}}
\def\mn@eprint@dblp#1{\href {http://dblp.uni-trier.de/rec/bibtex/#1.xml}
  {dblp:#1}}
\def\mn@eprint@#1:#2:#3:#4\@nil{\def\@tempa {#1}\def\@tempb {#2}\def\@tempc
  {#3}\ifx \@tempc \@empty \let \@tempc \@tempb \let \@tempb \@tempa \fi \ifx
  \@tempb \@empty \def\@tempb {arXiv}\fi \@ifundefined
  {mn@eprint@\@tempb}{\@tempb:\@tempc}{\expandafter \expandafter \csname
  mn@eprint@\@tempb\endcsname \expandafter{\@tempc}}}

\bibitem[\protect\citeauthoryear{{Annala} \& {Poutanen}}{{Annala} \&
  {Poutanen}}{2010}]{annala10}
{Annala} M.,  {Poutanen} J.,  2010, \mn@doi [\aap]
  {10.1051/0004-6361/200912773}, \href
  {https://ui.adsabs.harvard.edu/abs/2010A&A...520A..76A} {520, A76}

\bibitem[\protect\citeauthoryear{{Archibald} et~al.,}{{Archibald}
  et~al.}{2009}]{archibald09}
{Archibald} A.~M.,  et~al., 2009, \mn@doi [Science] {10.1126/science.1172740},
  \href {https://ui.adsabs.harvard.edu/abs/2009Sci...324.1411A} {324, 1411}

\bibitem[\protect\citeauthoryear{{Beskin} \& {Zheltoukhov}}{{Beskin} \&
  {Zheltoukhov}}{2014}]{bz_anomal2014}
{Beskin} V.~S.,  {Zheltoukhov} A.~A.,  2014, \mn@doi [Physics Uspekhi]
  {10.3367/UFNe.0184.201408e.0865}, \href
  {https://ui.adsabs.harvard.edu/abs/2014PhyU...57..799B} {57, 799}

\bibitem[\protect\citeauthoryear{{Beskin}, {Gurevich}  \& {Istomin}}{{Beskin}
  et~al.}{1993}]{bes93}
{Beskin} V.~S.,  {Gurevich} A.~V.,   {Istomin} Y.~N.,  1993, {Physics of the
  pulsar magnetosphere}.
{Cambridge, New York: Cambridge University Press}

\bibitem[\protect\citeauthoryear{{Bhattacharya} \& {van den
  Heuvel}}{{Bhattacharya} \& {van den Heuvel}}{1991}]{bhatt91}
{Bhattacharya} D.,  {van den Heuvel} E.~P.~J.,  1991, \mn@doi [\physrep]
  {10.1016/0370-1573(91)90064-S}, \href
  {https://ui.adsabs.harvard.edu/abs/1991PhR...203....1B} {203, 1}

\bibitem[\protect\citeauthoryear{{Biryukov}, {Astashenok}  \&
  {Beskin}}{{Biryukov} et~al.}{2017a}]{bab17}
{Biryukov} A.,  {Astashenok} A.,   {Beskin} G.,  2017a, \mn@doi [MNRAS]
  {10.1093/mnras/stw3341}, 466, 4320

\bibitem[\protect\citeauthoryear{{Biryukov}, {Astashenok}, {Karpov}  \&
  {Beskin}}{{Biryukov} et~al.}{2017b}]{bakb17}
{Biryukov} A.,  {Astashenok} A.,  {Karpov} S.,   {Beskin} G.,  2017b, in
  Journal of Physics Conference Series. p. 012044 (\mn@eprint {arXiv}
  {1711.05081}), \mn@doi{10.1088/1742-6596/932/1/012044}

\bibitem[\protect\citeauthoryear{{Bogdanov}, {Grindlay}  \&
  {Rybicki}}{{Bogdanov} et~al.}{2008}]{bogdanov08}
{Bogdanov} S.,  {Grindlay} J.~E.,   {Rybicki} G.~B.,  2008, \mn@doi [\apj]
  {10.1086/592341}, \href
  {https://ui.adsabs.harvard.edu/abs/2008ApJ...689..407B} {689, 407}

\bibitem[\protect\citeauthoryear{{Bozzo}, {Ascenzi}, {Ducci}, {Papitto},
  {Burderi}  \& {Stella}}{{Bozzo} et~al.}{2018}]{Bozzo18}
{Bozzo} E.,  {Ascenzi} S.,  {Ducci} L.,  {Papitto} A.,  {Burderi} L.,
  {Stella} L.,  2018, \mn@doi [\aap] {10.1051/0004-6361/201732004}, \href
  {https://ui.adsabs.harvard.edu/abs/2018A&A...617A.126B} {617, A126}

\bibitem[\protect\citeauthoryear{{Caballero} \& {Wilms}}{{Caballero} \&
  {Wilms}}{2012}]{CW12}
{Caballero} I.,  {Wilms} J.,  2012, \memsai, \href
  {https://ui.adsabs.harvard.edu/abs/2012MmSAI..83..230C} {83, 230}

\bibitem[\protect\citeauthoryear{{Carpano}, {Haberl}, {Maitra}  \&
  {Vasilopoulos}}{{Carpano} et~al.}{2018}]{carpano18}
{Carpano} S.,  {Haberl} F.,  {Maitra} C.,   {Vasilopoulos} G.,  2018, \mn@doi
  [\mnras] {10.1093/mnrasl/sly030}, \href
  {https://ui.adsabs.harvard.edu/abs/2018MNRAS.476L..45C} {476, L45}

\bibitem[\protect\citeauthoryear{{Casini} \& {Montemayor}}{{Casini} \&
  {Montemayor}}{1998}]{Casini1998}
{Casini} H.,  {Montemayor} R.,  1998, \mn@doi [\apj] {10.1086/305991}, \href
  {https://ui.adsabs.harvard.edu/abs/1998ApJ...503..374C} {503, 374}

\bibitem[\protect\citeauthoryear{{Chaty}}{{Chaty}}{2011}]{chaty11}
{Chaty} S.,  2011, in {Schmidtobreick} L.,  {Schreiber} M.~R.,   {Tappert} C.,
  eds,  Astronomical Society of the Pacific Conference Series Vol. 447,
  Evolution of Compact Binaries. p.~29 (\mn@eprint {arXiv} {1107.0231})

\bibitem[\protect\citeauthoryear{{Chen}, {Ruderman}  \& {Zhu}}{{Chen}
  et~al.}{1998}]{chen98}
{Chen} K.,  {Ruderman} M.,   {Zhu} T.,  1998, \mn@doi [\apj] {10.1086/305106},
  \href {https://ui.adsabs.harvard.edu/abs/1998ApJ...493..397C} {493, 397}

\bibitem[\protect\citeauthoryear{{Cheng}, {Zhang}, {Zhao}, {Wang}, {Pan}  \&
  {Lei}}{{Cheng} et~al.}{2014}]{Cheng2014}
{Cheng} Z.,  {Zhang} C.-M.,  {Zhao} Y.-H.,  {Wang} D.-H.,  {Pan} Y.-Y.,   {Lei}
  Y.-J.,  2014, \mn@doi [\caa] {10.1016/j.chinastron.2014.07.006}, \href
  {https://ui.adsabs.harvard.edu/abs/2014ChA&A..38..294C} {38, 294}

\bibitem[\protect\citeauthoryear{{Contopoulos} \& {Spitkovsky}}{{Contopoulos}
  \& {Spitkovsky}}{2006}]{cs06}
{Contopoulos} I.,  {Spitkovsky} A.,  2006, \mn@doi [\apj] {10.1086/501161},
  \href {https://ui.adsabs.harvard.edu/abs/2006ApJ...643.1139C} {643, 1139}

\bibitem[\protect\citeauthoryear{{Frank}, {King}  \& {Raine}}{{Frank}
  et~al.}{2002}]{accpower}
{Frank} J.,  {King} A.,   {Raine} D.~J.,  2002, {Accretion Power in
  Astrophysics: Third Edition}

\bibitem[\protect\citeauthoryear{{Fryxell} \& {Taam}}{{Fryxell} \&
  {Taam}}{1988}]{FT88}
{Fryxell} B.~A.,  {Taam} R.~E.,  1988, \mn@doi [\apj] {10.1086/166973}, \href
  {http://adsabs.harvard.edu/abs/1988ApJ...335..862F} {335, 862}

\bibitem[\protect\citeauthoryear{{Ghosh}, {Lamb}  \& {Pethick}}{{Ghosh}
  et~al.}{1977}]{GL77}
{Ghosh} P.,  {Lamb} F.~K.,   {Pethick} C.~J.,  1977, \mn@doi [\apj]
  {10.1086/155606}, \href
  {https://ui.adsabs.harvard.edu/abs/1977ApJ...217..578G} {217, 578}

\bibitem[\protect\citeauthoryear{{Goldreich}}{{Goldreich}}{1970}]{goldreich70}
{Goldreich} P.,  1970, \mn@doi [\apjl] {10.1086/180513}, \href
  {https://ui.adsabs.harvard.edu/abs/1970ApJ...160L..11G} {160, L11}

\bibitem[\protect\citeauthoryear{{Gull{\'o}n}, {Miralles}, {Vigan{\`o}}  \&
  {Pons}}{{Gull{\'o}n} et~al.}{2014}]{gullon14}
{Gull{\'o}n} M.,  {Miralles} J.~A.,  {Vigan{\`o}} D.,   {Pons} J.~A.,  2014,
  \mn@doi [MNRAS] {10.1093/mnras/stu1253}, \href
  {http://adsabs.harvard.edu/abs/2014MNRAS.443.1891G} {443, 1891}

\bibitem[\protect\citeauthoryear{{Haskell}, {Samuelsson}, {Glampedakis}  \&
  {Andersson}}{{Haskell} et~al.}{2008}]{haskell08}
{Haskell} B.,  {Samuelsson} L.,  {Glampedakis} K.,   {Andersson} N.,  2008,
  \mn@doi [\mnras] {10.1111/j.1365-2966.2008.12861.x}, \href
  {http://adsabs.harvard.edu/abs/2008MNRAS.385..531H} {385, 531}

\bibitem[\protect\citeauthoryear{{Hessels}}{{Hessels}}{2008}]{hessels08}
{Hessels} J.~W.~T.,  2008, in {Wijnands} R.,  {Altamirano} D.,  {Soleri} P.,
  {Degenaar} N.,  {Rea} N.,  {Casella} P.,  {Patruno} A.,   {Linares} M.,  eds,
   American Institute of Physics Conference Series Vol. 1068, American
  Institute of Physics Conference Series. pp 130--134 (\mn@eprint {arXiv}
  {0903.0493}), \mn@doi{10.1063/1.3031183}

\bibitem[\protect\citeauthoryear{{Igoshev} \& {Popov}}{{Igoshev} \&
  {Popov}}{2015}]{ip15}
{Igoshev} A.~P.,  {Popov} S.~B.,  2015, \mn@doi [Astronomische Nachrichten]
  {10.1002/asna.201512232}, \href
  {http://adsabs.harvard.edu/abs/2015AN....336..831I} {336, 831}

\bibitem[\protect\citeauthoryear{{Illarionov} \& {Kompaneets}}{{Illarionov} \&
  {Kompaneets}}{1990}]{IK90}
{Illarionov} A.~F.,  {Kompaneets} D.~A.,  1990, \mnras, \href
  {http://adsabs.harvard.edu/abs/1990MNRAS.247..219I} {247, 219}

\bibitem[\protect\citeauthoryear{{Illarionov} \& {Sunyaev}}{{Illarionov} \&
  {Sunyaev}}{1975}]{IS75}
{Illarionov} A.~F.,  {Sunyaev} R.~A.,  1975, \aap, \href
  {https://ui.adsabs.harvard.edu/abs/1975A&A....39..185I} {39, 185}

\bibitem[\protect\citeauthoryear{{Johnson} et~al.,}{{Johnson}
  et~al.}{2014}]{johnson14}
{Johnson} T.~J.,  et~al., 2014, \mn@doi [\apjs] {10.1088/0067-0049/213/1/6},
  \href {https://ui.adsabs.harvard.edu/abs/2014ApJS..213....6J} {213, 6}

\bibitem[\protect\citeauthoryear{{Lai}}{{Lai}}{2014}]{lai2014}
{Lai} D.,  2014, in European Physical Journal Web of Conferences. p. 01001
  (\mn@eprint {arXiv} {1402.1903}), \mn@doi{10.1051/epjconf/20136401001}

\bibitem[\protect\citeauthoryear{{Lai}, {Foucart}  \& {Lin}}{{Lai}
  et~al.}{2011}]{lai11}
{Lai} D.,  {Foucart} F.,   {Lin} D. N.~C.,  2011, \mn@doi [\mnras]
  {10.1111/j.1365-2966.2010.18127.x}, \href
  {https://ui.adsabs.harvard.edu/abs/2011MNRAS.412.2790L} {412, 2790}

\bibitem[\protect\citeauthoryear{Landau, Lifshitz, Sykes  \& Bell}{Landau
  et~al.}{1976}]{LL1}
Landau L.,  Lifshitz E.,  Sykes J.,   Bell J.,  1976, Mechanics.
Butterworth-Heinemann, Elsevier Science, \url
  {https://books.google.fi/books?id=e-xASAehg1sC}

\bibitem[\protect\citeauthoryear{{Leahy}}{{Leahy}}{1990}]{leahy90}
{Leahy} D.~A.,  1990, \mn@doi [\mnras] {10.1093/mnras/242.2.188}, \href
  {https://ui.adsabs.harvard.edu/abs/1990MNRAS.242..188L} {242, 188}

\bibitem[\protect\citeauthoryear{{Lipunov} \& {Shakura}}{{Lipunov} \&
  {Shakura}}{1980}]{LS80}
{Lipunov} V.~M.,  {Shakura} N.~I.,  1980, Soviet Astronomy Letters, \href
  {https://ui.adsabs.harvard.edu/abs/1980SvAL....6...14L} {6, 14}

\bibitem[\protect\citeauthoryear{{Lovelace}, {Romanova}  \&
  {Bisnovatyi-Kogan}}{{Lovelace} et~al.}{1995}]{lovelace95}
{Lovelace} R.~V.~E.,  {Romanova} M.~M.,   {Bisnovatyi-Kogan} G.~S.,  1995,
  \mn@doi [\mnras] {10.1093/mnras/275.2.244}, \href
  {http://adsabs.harvard.edu/abs/1995MNRAS.275..244L} {275, 244}

\bibitem[\protect\citeauthoryear{{Lyne} \& {Manchester}}{{Lyne} \&
  {Manchester}}{1988}]{lyne88}
{Lyne} A.~G.,  {Manchester} R.~N.,  1988, \mn@doi [MNRAS]
  {10.1093/mnras/234.3.477}, \href
  {http://adsabs.harvard.edu/abs/1988MNRAS.234..477L} {234, 477}

\bibitem[\protect\citeauthoryear{{MacLeod} \& {Ramirez-Ruiz}}{{MacLeod} \&
  {Ramirez-Ruiz}}{2015}]{MMRR}
{MacLeod} M.,  {Ramirez-Ruiz} E.,  2015, \mn@doi [\apj]
  {10.1088/0004-637X/803/1/41}, \href
  {https://ui.adsabs.harvard.edu/abs/2015ApJ...803...41M} {803, 41}

\bibitem[\protect\citeauthoryear{{Malov} \& {Nikitina}}{{Malov} \&
  {Nikitina}}{2011}]{nikitina11}
{Malov} I.~F.,  {Nikitina} E.~B.,  2011, \mn@doi [Astronomy Reports]
  {10.1134/S1063772911100076}, \href
  {http://adsabs.harvard.edu/abs/2011ARep...55..878M} {55, 878}

\bibitem[\protect\citeauthoryear{{Manchester}}{{Manchester}}{2017}]{manchester_msp_2017}
{Manchester} R.~N.,  2017, \mn@doi [Journal of Astrophysics and Astronomy]
  {10.1007/s12036-017-9469-2}, \href
  {https://ui.adsabs.harvard.edu/abs/2017JApA...38...42M} {38, 42}

\bibitem[\protect\citeauthoryear{{Melatos}}{{Melatos}}{2000}]{melatos2000}
{Melatos} A.,  2000, \mn@doi [MNRAS] {10.1046/j.1365-8711.2000.03031.x}, \href
  {http://adsabs.harvard.edu/abs/2000MNRAS.313..217M} {313, 217}

\bibitem[\protect\citeauthoryear{{Melatos} \& {Mastrano}}{{Melatos} \&
  {Mastrano}}{2016}]{Melatos2016_Poison}
{Melatos} A.,  {Mastrano} A.,  2016, \mn@doi [\apj]
  {10.3847/0004-637X/818/1/49}, \href
  {https://ui.adsabs.harvard.edu/abs/2016ApJ...818...49M} {818, 49}

\bibitem[\protect\citeauthoryear{{Melatos} \& {Phinney}}{{Melatos} \&
  {Phinney}}{2001}]{MP01}
{Melatos} A.,  {Phinney} E.~S.,  2001, \mn@doi [Publications of the
  Astronomical Society of Australia] {10.1071/AS01056}, \href
  {https://ui.adsabs.harvard.edu/#abs/2001PASA...18..421M} {18, 421}

\bibitem[\protect\citeauthoryear{{Novoselov}, {Beskin}, {Galishnikova},
  {Rashkovetskyi}  \& {Biryukov}}{{Novoselov} et~al.}{2020}]{Novoselov2020}
{Novoselov} E.~M.,  {Beskin} V.~S.,  {Galishnikova} A.~K.,  {Rashkovetskyi}
  M.~M.,   {Biryukov} A.~V.,  2020, \mn@doi [\mnras] {10.1093/mnras/staa904},
  \href {https://ui.adsabs.harvard.edu/abs/2020MNRAS.494.3899N} {494, 3899}

\bibitem[\protect\citeauthoryear{{Pan}, {Zhang}  \& {Wang}}{{Pan}
  et~al.}{2013}]{pan2013}
{Pan} Y.,  {Zhang} C.,   {Wang} N.,  2013, in {Zhang} C.~M.,  {Belloni} T.,
  {M{\'e}ndez} M.,   {Zhang} S.~N.,  eds,  Proceedings of the International
  Astronomical Union Vol. 290, Feeding Compact Objects: Accretion on All
  Scales. pp 291--292 (\mn@eprint {arXiv} {1304.0073}),
  \mn@doi{10.1017/S1743921312020066}

\bibitem[\protect\citeauthoryear{{Papitto} \& {de Martino}}{{Papitto} \& {de
  Martino}}{2020}]{Papitto2020_TPSR}
{Papitto} A.,  {de Martino} D.,  2020, arXiv e-prints, \href
  {https://ui.adsabs.harvard.edu/abs/2020arXiv201009060P} {p. arXiv:2010.09060}

\bibitem[\protect\citeauthoryear{{Parfrey}, {Spitkovsky}  \&
  {Beloborodov}}{{Parfrey} et~al.}{2016}]{parfrey16}
{Parfrey} K.,  {Spitkovsky} A.,   {Beloborodov} A.~M.,  2016, \mn@doi [\apj]
  {10.3847/0004-637X/822/1/33}, \href
  {https://ui.adsabs.harvard.edu/#abs/2016ApJ...822...33P} {822, 33}

\bibitem[\protect\citeauthoryear{{Parfrey}, {Spitkovsky}  \&
  {Beloborodov}}{{Parfrey} et~al.}{2017}]{parfrey17}
{Parfrey} K.,  {Spitkovsky} A.,   {Beloborodov} A.~M.,  2017, \mn@doi [\mnras]
  {10.1093/mnras/stx950}, \href
  {https://ui.adsabs.harvard.edu/abs/2017MNRAS.469.3656P} {469, 3656}

\bibitem[\protect\citeauthoryear{{Patruno} \& {Watts}}{{Patruno} \&
  {Watts}}{2021}]{AMSP21}
{Patruno} A.,  {Watts} A.~L.,  2021, \mn@doi [Astrophysics and Space Science
  Library] {10.1007/978-3-662-62110-3_4}, \href
  {https://ui.adsabs.harvard.edu/abs/2021ASSL..461..143P} {461, 143}

\bibitem[\protect\citeauthoryear{{P{\'e}tri}}{{P{\'e}tri}}{2019}]{Petri2019}
{P{\'e}tri} J.,  2019, \mn@doi [\mnras] {10.1093/mnras/stz711}, \href
  {https://ui.adsabs.harvard.edu/abs/2019MNRAS.485.4573P} {485, 4573}

\bibitem[\protect\citeauthoryear{{Philippov}, {Tchekhovskoy}  \&
  {Li}}{{Philippov} et~al.}{2014}]{phil14}
{Philippov} A.,  {Tchekhovskoy} A.,   {Li} J.~G.,  2014, \mn@doi [MNRAS]
  {10.1093/mnras/stu591}, \href
  {http://adsabs.harvard.edu/abs/2014MNRAS.441.1879P} {441, 1879}

\bibitem[\protect\citeauthoryear{{Rankin}}{{Rankin}}{1993}]{rankin93b}
{Rankin} J.~M.,  1993, \mn@doi [Astrophys. J.] {10.1086/172361}, \href
  {http://adsabs.harvard.edu/abs/1993ApJ...405..285R} {405, 285}

\bibitem[\protect\citeauthoryear{{Rappaport}, {Fregeau}  \&
  {Spruit}}{{Rappaport} et~al.}{2004}]{Rap2004}
{Rappaport} S.~A.,  {Fregeau} J.~M.,   {Spruit} H.,  2004, \mn@doi [\apj]
  {10.1086/382863}, \href
  {https://ui.adsabs.harvard.edu/abs/2004ApJ...606..436R} {606, 436}

\bibitem[\protect\citeauthoryear{{Reisenegger}}{{Reisenegger}}{2003}]{reis03}
{Reisenegger} A.,  2003, ArXiv Astrophysics e-prints, \href
  {http://adsabs.harvard.edu/abs/2003astro.ph..7133R} {}

\bibitem[\protect\citeauthoryear{{Romanova}, {Ustyugova}, {Koldoba}  \&
  {Lovelace}}{{Romanova} et~al.}{2009}]{romanova09}
{Romanova} M.~M.,  {Ustyugova} G.~V.,  {Koldoba} A.~V.,   {Lovelace} R.~V.~E.,
  2009, \mn@doi [\mnras] {10.1111/j.1365-2966.2009.15413.x}, \href
  {http://adsabs.harvard.edu/abs/2009MNRAS.399.1802R} {399, 1802}

\bibitem[\protect\citeauthoryear{{Romanova}, {Koldoba}, {Ustyugova}, {Blinova},
  {Lai}  \& {Lovelace}}{{Romanova} et~al.}{2020}]{romanova20}
{Romanova} M.~M.,  {Koldoba} A.~V.,  {Ustyugova} G.~V.,  {Blinova} A.~A.,
  {Lai} D.,   {Lovelace} R.~V.~E.,  2020, arXiv e-prints, \href
  {https://ui.adsabs.harvard.edu/abs/2020arXiv201210826R} {p. arXiv:2012.10826}

\bibitem[\protect\citeauthoryear{{Ruderman}}{{Ruderman}}{1991}]{Ruderman1991}
{Ruderman} R.,  1991, \mn@doi [\apj] {10.1086/170744}, \href
  {https://ui.adsabs.harvard.edu/abs/1991ApJ...382..576R} {382, 576}

\bibitem[\protect\citeauthoryear{{Ruffert}}{{Ruffert}}{1997}]{ruffert97}
{Ruffert} M.,  1997, \aap, \href
  {http://adsabs.harvard.edu/abs/1997A%26A...317..793R} {317, 793}

\bibitem[\protect\citeauthoryear{{Ruffert}}{{Ruffert}}{1999}]{ruffert99}
{Ruffert} M.,  1999, \aap, \href
  {http://adsabs.harvard.edu/abs/1999A%26A...346..861R} {346, 861}

\bibitem[\protect\citeauthoryear{{Shakura}, {Postnov}, {Kochetkova}  \&
  {Hjalmarsdotter}}{{Shakura} et~al.}{2012}]{shakura12}
{Shakura} N.,  {Postnov} K.,  {Kochetkova} A.,   {Hjalmarsdotter} L.,  2012,
  \mn@doi [\mnras] {10.1111/j.1365-2966.2011.20026.x}, \href
  {http://adsabs.harvard.edu/abs/2012MNRAS.420..216S} {420, 216}

\bibitem[\protect\citeauthoryear{{Shvartsman}}{{Shvartsman}}{1970}]{shvartsman70}
{Shvartsman} V.~F.,  1970, Izvestiia Vysshaia Uchebn.~Zaved., Radiofizika,
  \href {http://adsabs.harvard.edu/abs/1970IzVUZ..13.1852S} {13, 1852}

\bibitem[\protect\citeauthoryear{{Shvartsman}}{{Shvartsman}}{1971}]{Shvartsman1971}
{Shvartsman} V.~F.,  1971, \sovast, \href
  {https://ui.adsabs.harvard.edu/abs/1971SvA....15..342S} {15, 342}

\bibitem[\protect\citeauthoryear{{Spitkovsky}}{{Spitkovsky}}{2006}]{spitkovsky06}
{Spitkovsky} A.,  2006, \mn@doi [Astrophys. J. Lett.] {10.1086/507518}, \href
  {http://adsabs.harvard.edu/abs/2006ApJ...648L..51S} {648, L51}

\bibitem[\protect\citeauthoryear{{Tauris} \& {Manchester}}{{Tauris} \&
  {Manchester}}{1998}]{tm98}
{Tauris} T.~M.,  {Manchester} R.~N.,  1998, \mn@doi [MNRAS]
  {10.1046/j.1365-8711.1998.01369.x}, \href
  {http://adsabs.harvard.edu/abs/1998MNRAS.298..625T} {298, 625}

\bibitem[\protect\citeauthoryear{{Tauris} \& {van den Heuvel}}{{Tauris} \& {van
  den Heuvel}}{2006}]{taurisheuvel}
{Tauris} T.~M.,  {van den Heuvel} E.~P.~J.,  2006, {Formation and evolution of
  compact stellar X-ray sources}.
pp 623--665

\bibitem[\protect\citeauthoryear{{Tauris}, {Langer}  \& {Kramer}}{{Tauris}
  et~al.}{2012}]{tauris2012}
{Tauris} T.~M.,  {Langer} N.,   {Kramer} M.,  2012, \mn@doi [\mnras]
  {10.1111/j.1365-2966.2012.21446.x}, \href
  {https://ui.adsabs.harvard.edu/abs/2012MNRAS.425.1601T} {425, 1601}

\bibitem[\protect\citeauthoryear{{Vasilopoulos}, {Haberl}, {Carpano}  \&
  {Maitra}}{{Vasilopoulos} et~al.}{2018}]{vasilopoulos}
{Vasilopoulos} G.,  {Haberl} F.,  {Carpano} S.,   {Maitra} C.,  2018, \mn@doi
  [\aap] {10.1051/0004-6361/201833442}, \href
  {https://ui.adsabs.harvard.edu/\#abs/2018A&A...620L..12V} {620, L12}

\bibitem[\protect\citeauthoryear{{Veledina}, {N{\"a}ttil{\"a}}  \&
  {Beloborodov}}{{Veledina} et~al.}{2019}]{veledina19}
{Veledina} A.,  {N{\"a}ttil{\"a}} J.,   {Beloborodov} A.~M.,  2019, \mn@doi
  [\apj] {10.3847/1538-4357/ab44c6}, \href
  {https://ui.adsabs.harvard.edu/abs/2019ApJ...884..144V} {884, 144}

\bibitem[\protect\citeauthoryear{{Wang}}{{Wang}}{1981a}]{wang_inclined}
{Wang} Y.~M.,  1981a, \mn@doi [\ssr] {10.1007/BF01246043}, \href
  {https://ui.adsabs.harvard.edu/abs/1981SSRv...30..341W} {30, 341}

\bibitem[\protect\citeauthoryear{{Wang}}{{Wang}}{1981b}]{wang81}
{Wang} Y.-M.,  1981b, \aap, \href
  {http://adsabs.harvard.edu/abs/1981A%26A...102...36W} {102, 36}

\bibitem[\protect\citeauthoryear{{Zanni} \& {Ferreira}}{{Zanni} \&
  {Ferreira}}{2013}]{ZF13}
{Zanni} C.,  {Ferreira} J.,  2013, \mn@doi [\aap]
  {10.1051/0004-6361/201220168}, \href
  {http://adsabs.harvard.edu/abs/2013A%26A...550A..99Z} {550, A99}

\bibitem[\protect\citeauthoryear{{Zhang} et~al.,}{{Zhang}
  et~al.}{2011}]{Zhang2011}
{Zhang} C.~M.,  et~al., 2011, \mn@doi [\aap] {10.1051/0004-6361/201015532},
  \href {https://ui.adsabs.harvard.edu/abs/2011A&A...527A..83Z} {527, A83}

\makeatother
\end{thebibliography}

\appendix 

\section{Spheroidal neutron star}\label{sec:app}

Let us consider a spheroidal NS having variable ellipticity $\epsilon$ (positive or negative) along an arbitrary axis $\mathbi{l}$ ($|\mathbi{l}| = 1$). 
Then the equation for the rotational evolution of such a NS may be written as 
\begin{equation}
    I \dfrac{\diff\mathbi{\Omega}}{\diff \it t} = \mathbi{N} - \dfrac{\diff \it I}{\diff \it t} \mathbi{\Omega}
    -\epsilon I \Omega_{l} \left [ \left (\dfrac{\dot I}{I} + \dfrac{\dot\Omega_l}{\Omega_l}  + \dfrac{\dot\epsilon}{\epsilon }\right ) \mathbi{l} + \mathbi{\Omega} \times \mathbi {l} \right],
	\label{eq:general_biaxial_eqs}
\end{equation}
where $\Omega_l$ is the scalar product $\mathbi{\Omega}\cdot \mathbi{l}$.

The right-hand side of Equation~(\ref{eq:general_biaxial_eqs}) consists of three terms: external torque $\mathbi{N}$, the term proportional to $\diff {\it I}/\diff {\it t}$ and related to the variable moment of inertia 
of the star, and the term arising from non-sphericity. The latter
can be also rewritten as
\begin{equation}\label{E:Ne}
    {\mathbi N}_{\epsilon} = -(\epsilon\Omega_l\dot I + \epsilon I \dot \Omega_l + \dot \epsilon I \Omega_l) \mathbi{l} + \epsilon I \Omega_l (\mathbi{\Omega} \times \mathbi{l}).
\end{equation}
Since the ellipticity of a neutron star is expected to be very small (i.e. $|\epsilon| \ll 1$, see for instance~\citealt{haskell08}) and changes slowly, 
\begin{equation}
    |\epsilon \Omega_l \dot I| \ll |\Omega \dot I|
\end{equation}
and
\begin{equation}
    |\epsilon I \dot \Omega_l| \ll |I\dot \Omega| \simeq I\Omega\tau^{-1}_\mathrm{align},
\end{equation}
as the alignment time $\tau_\mathrm{align}$ and the spin-up time scale are determined by the same external torque $\mathbi{N}$.

Therefore, the first term in $\mathbi{N}_{\epsilon}$, aligned with the axis $\mathbi{l}$, may be omitted while considering the evolution of an accreting neutron star during spin-disc alignment, as the contribution of the alignment itself is always larger. 

If the deformation is aligned with the magnetic moment (or, equivalently, $\mathbi{l} = \mathbi{m}$), the second term in Equation~(\ref{E:Ne}), proportional to $\mathbi{\Omega}\times \mathbi{l}$, does not contribute to the evolutionary equations for the reasons discussed in Section~\ref{sec:modulation} for the anomalous pulsar torque and valid for any torque orthogonal to the magnetic and spin axes at the same time.

The only possible case when the precession term may become important for the long-term evolution of the rotation parameters is the case when $\mathbi{l}$ is not co-planar with the spin and magnetic axes. 

But this situation requires a triaxial rather than a spheroidal star \citep{melatos2000}, which is beyond of the scope of current paper.

\label{lastpage}

\end{document}